\documentclass[twocolumn,trackchanges]{aastex63}
\usepackage{xspace,color,soul}

\newcommand{\per}{\ensuremath{^{-1}}\xspace}
\newcommand{\Lya}{Ly\ensuremath{\alpha}\xspace}

\received{March 6, 2020}
\revised{April 17, 2020}
\accepted{April 18, 2020}

\submitjournal{\apj}

\shorttitle{The Ultraluminous Ly$\alpha$ Luminosity Function at $z=6.6$}
\shortauthors{Taylor et al.}

\begin{document}

\title{The Ultraluminous Ly$\alpha$ Luminosity Function at $z=6.6$}

\correspondingauthor{Anthony J. Taylor}
\email{ataylor@astro.wisc.edu}

\author[0000-0003-1282-7454]{A.~J.~Taylor}
\affiliation{Department of Astronomy, University of Wisconsin-Madison,
475 N. Charter Street, Madison, WI 53706, USA}

\author[0000-0002-3306-1606]{A.~J.~Barger}
\affiliation{Department of Astronomy, University of Wisconsin-Madison,
475 N. Charter Street, Madison, WI 53706, USA}
\affiliation{Department of Physics and Astronomy, University of Hawaii,
2505 Correa Road, Honolulu, HI 96822, USA}
\affiliation{Institute for Astronomy, University of Hawaii, 2680 Woodlawn Drive,
Honolulu, HI 96822, USA}

\author[0000-0002-6319-1575]{L.~L.~Cowie}
\affiliation{Institute for Astronomy, University of Hawaii,
2680 Woodlawn Drive, Honolulu, HI 96822, USA}

\author{E.~M.~Hu}
\affiliation{Institute for Astronomy, University of Hawaii,
2680 Woodlawn Drive, Honolulu, HI 96822, USA}

\author{A.~Songaila}
\affiliation{Institute for Astronomy, University of Hawaii,
2680 Woodlawn Drive, Honolulu, HI 96822, USA}

\begin{abstract}
We present the luminosity function (LF) for ultraluminous \Lya emitting galaxies (LAEs) 
at $z=6.6$. 
We define ultraluminous LAEs (ULLAEs) as galaxies with 
$\log L (\textrm{Ly}\alpha)>43.5$~erg~s\per.
We select our main sample using the
$g'$, $r'$, $i'$, $z'$, and NB921 observations of
a wide-area (30~deg$^2$) Hyper Suprime-Cam
survey of the North Ecliptic Pole (NEP) field.
We select candidates with $g',r',i'>26$, NB$921\leq23.5$, and 
NB$921-z'\leq1.3$. Using the DEIMOS spectrograph on Keck~II,
we confirm 9 of our 14 candidates as ULLAEs at $z=6.6$ 
and the remaining 5 as
an AGN at $z=6.6$, two [OIII]$\lambda$5007 emitting galaxies at $z=0.84$ 
and $z=0.85$, and two non-detections. 
This emphasizes the need for full spectroscopic follow-up to determine
accurate LFs. In constructing the ULLAE LF at $z=6.6$, we combine our 9 NEP
ULLAEs with two previously discovered and confirmed ULLAEs in the COSMOS 
field:  CR7 and COLA1. We apply rigorous corrections for incompleteness based on simulations.
We compare our ULLAE LF at $z=6.6$ with LFs at $z=5.7$ and $z=6.6$ 
from the literature. Our data reject some previous LF normalizations and power law indices,
but they are broadly consistent with others. Indeed, a comparative analysis
of the different literature LFs suggests that none is fully consistent 
with any of the others, making it critical to 
determine the evolution from $z=5.7$ 
to $z=6.6$ using LFs constructed in exactly the same way at both redshifts.
\end{abstract}

\section{Introduction}
The epoch of reionization, in which the intergalactic medium (IGM) transitioned from 
being dominantly neutral to being primarily ionized, is a key era in the universe's history.  
\Lya emission from early, massively star-forming galaxies is one of the few probes of 
galaxy evolution and activity in this era. This early \Lya emission is redshifted from the 
rest-frame UV into the observed-frame optical, whereas other less energetic diagnostic 
emission lines are redshifted into the more difficult to observe infrared. 

Narrowband ($\sim100$~\AA) surveys can identify \Lya emitter (LAE) candidates 
at specific redshifts by imaging low-background windows in the atmospheric sky. 
In recent years, narrowband surveys have identified LAE candidates 
out to redshifts beyond $z=7$ 
\citep[e.g.,][]{hu04,hu10,hu16,ouchi08,ouchi10,kashikawa11,konno14,konno18,matthee15,santos16,jaing17,ota17,songaila18,itoh18,LAGER19}.  
With this large influx of new samples, the very high-redshift \Lya luminosity function 
(LF) is being probed. However, most of these LFs suffer from 
low number statistics and large errors, especially at the ultraluminous 
($\log L (\textrm{Ly}\alpha)>43.5$~erg~s\per) end. 
The rarity of ultraluminous LAEs (ULLAEs) makes the population highly 
susceptible to contamination from foreground strong emission line galaxies 
and from active galactic nuclei (AGNs), which means spectroscopic confirmation
is critical. Additionally, wide-area surveys are needed to reduce possible effects of cosmic variance. 

The evolution of the LAE LF potentially offers insight into the onset of reionization.  
In particular, \cite{santos16} have claimed to see differential evolution in the shape of the 
LAE LF at high redshift, i.e., a significant decline in the number density of “typical” LAEs 
from $z=5.7$ to $z=6.6$, and no evolution of ULLAEs over the same redshift range.
Such a result might imply that reionization is being completed first around ULLAEs,
since the increasing neutrality would have a larger effect on the luminosities
of the lower luminosity LAEs than on those of the ULLAEs.
Moreover, it would be consistent with the interpretation of the discovery of some
complex line profiles for $z=6.6$ ULLAEs \citep{hu16,songaila18} as
possible evidence for ULLAEs generating highly ionized regions of the IGM in 
their vicinity, thereby allowing the full \Lya profile of the galaxy, including blue wings,
to become visible.

Suprime-Cam and Hyper Suprime-Cam (HSC) narrowband surveys
\citep[e.g.,][]{ouchi08,ouchi10,hu10,kashikawa11,matthee15,santos16,zheng17,konno18}
are not all in agreement on the evolution of the LFs, 
though generally a decrease from $z=5.7$ to $z=6.6$
for non-ULLAEs is claimed. Most of these surveys 
\citep[other than the spectroscopically complete survey of][]{hu10}
have relatively few spectroscopic redshifts,
and the measured values at the ultraluminous end, when they exist at all, 
typically have large uncertainties.
Without spectroscopy, contaminants are likely to be included in the LAE samples,
which will result in higher normalizations for the LFs.  There is also the issue of
incompleteness corrections. For Suprime-Cam, the filters are 
far from top-hat. Thus, without knowing the exact redshift of a source, and hence 
where it lies in the filter, it is very difficult to correct the \Lya luminosity 
for the effect of the filter transmission profile. This will impact the number densities of bright 
LAEs, as some of these will be observed at a fraction of their luminosities.
In addition, some fainter LAEs will fall below the selection magnitude limit 
and hence not be included in the sample.
For HSC, the more top-hat like filters lessen these effects, but 
corrections are still needed. 

On the other hand, it was pointed out by \cite{kashikawa11} that
spectroscopically complete samples could 
result in lower normalizations for the LFs if insufficiently deep spectral data 
are used for the \Lya line identifications, since some 
candidates from the initial selection might be incorrectly removed.

The goal of the present paper is to construct the ULLAE portion of the LF 
at $z=6.6$ using a spectroscopically complete sample with sufficiently 
deep spectra that there is no ambiguity about the redshift identifications.
This prevents the sample from being contaminated by non-LAEs, and 
it also means that the \Lya luminosity corrections due to the filter transmission
profile are exact.

\citet{hu16} obtained narrowband NB921 images with HSC of the 
COSMOS field, and \citet{songaila18}
did the same for a much larger area in the North Ecliptic Pole (NEP) field.
From these data, we now have 11 spectroscopically confirmed ULLAEs at 
$z=6.6$ from which to construct the ultraluminous end of the LF.
We will then make comparisons of this new ULLAE LF
with LAE LFs at $z=5.7$ and $z=6.6$ from the literature. 

We assume 
$\Omega_{\rm M}$=0.3, $\Omega_{\Lambda}$=0.7, and H$_0$=70~km~s\per Mpc\per 
throughout. We give all magnitudes in the AB magnitude system, where an AB 
magnitude is defined by $m_{AB} = -2.5 \log f_{\nu}-48.60$. We define $f_{\nu}$, 
the flux of the source, in units of erg~cm$^{-2}$~s\per~Hz\per.

\section{Candidate Selection}
\begin{figure}[ht]
\begin{centering}
\includegraphics[width=\linewidth]{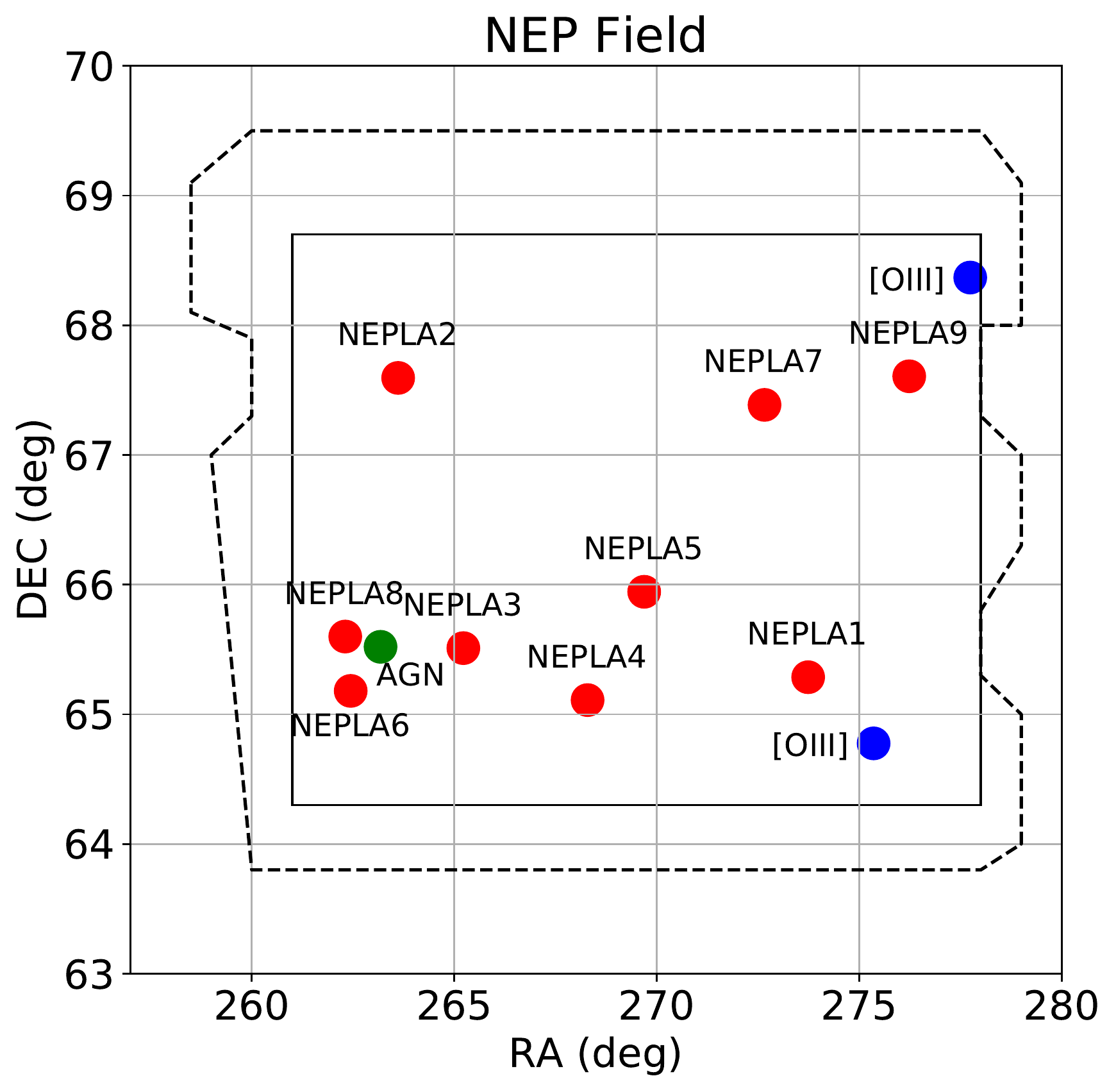}
\caption{
The dashed line shows the full $\sim$45 deg$^2$ of the current HEROES
survey, and the solid rectangle shows the most uniformly covered
30~deg$^2$ area studied in this work. 
Red circles denote the positions of the 9 spectroscopically confirmed LAEs, 
blue circles indicate the two [OIII] emitters, and the green circle 
indicates the single AGN. There were also two candidates that turned out to be 
spurious after spectroscopic observations.
}
\label{fig:NEPmap}
\end{centering}
\end{figure}

We use data from the \textit{H}awaii \textit{eRO}SITA \textit{E}cliptic Pole \textit{S}urvey,
or HEROES, a 45~deg$^2$ imaging survey of the North Ecliptic Pole (NEP;
dashed line in Figure~\ref{fig:NEPmap}; see \citealt{songaila18}).
HEROES consists of Subaru 8.2~m HSC broadband $g'$, $r'$, $i'$, $z'$, and 
$y'$ and narrowband NB816 and NB921 imaging.
(The $r'$ and $i'$ filters are the HSC-r2 and HSC-i2 filters.)
HEROES also includes
$U$ and $J$ imaging from the Canada-France-Hawaii Telescope (CFHT) 
3.6~m MegaPrime/MegaCam and WIRCam instruments.
In this work, we only use the $g'$, $r'$, $i'$, $z'$, and NB921 data, and
we focus on the most uniformly covered area, which is 30~deg$^2$ in size
(black rectangle in Figure~\ref{fig:NEPmap}).
For the NB921 data,
the $1\sigma$ noise in a $2''$ diameter aperture ranges from 25.4--26.1.
For the $z'$-band data, the $1\sigma$ noise in a $2''$ diameter aperture 
ranges from 26.2--27.0.
A detailed discussion of both the observations and the data reduction with
the Pan-STARRS Image Processing Pipeline \citep{magnier16} can be found in
\cite{songaila18}.

\cite{songaila18} selected objects with NB921 Kron magnitudes brighter than 23.5 in
the 30~deg$^2$ area. 
This selection in NB921 provides a $>5\sigma$ criterion throughout the area and a 
much higher signal-to-noise (mean of $9\sigma$) through most of the area.
This yielded a sample of 2.8 million objects.
For these objects, they centered on the NB921 positions and measured the magnitudes 
across the filters using $2''$ diameter apertures. They then restricted
to sources with $z'$--NB921~$\geq1.3$ that were also not detected above 
the 2$\sigma$ level in any of the $g'$, $r'$ and $i'$ bands, as would be
expected for LAEs with a strong Lyman break. 
The $z'$--NB921~$\geq1.3$ color excess selects high equivalent 
width spectra (observed-frame EW(\Lya)~$>100$~\AA) at redshifts $6.50<z<6.63$.
This choice of color excess selects all galaxies with rest-frame EW(\Lya)~$\ge20$~\AA, 
which is the normal definition of a high-redshift LAE \citep{hu98}.
Lyman continuum break sources at redshifts placing the break near the upper
wavelength of the narrowband filter can also satisfy these selection criteria, again
emphasizing the need for spectroscopic follow-up.
We summarize our selection criteria in Table~1.

We reanalyzed the selected candidates, looking for contamination 
from glints, bright stars, moving objects, etc. We then stacked the $g'$, $r'$, and $i'$ 
images for each of the remaining candidates and discarded any sources that appeared
in both the narrowband and stacked images. After our reanalysis, we ended up 
with 14 candidates for spectroscopic followup, 13 of which are in common with 
\cite{songaila18}.

\begin{deluxetable}{cc}
\renewcommand\baselinestretch{1.0}
\tablewidth{0pt}
\tablecaption{Photometric Selection Criteria}
\scriptsize
\tablehead{Filter & Selection}
\startdata
$g'$ & $>26$ \cr
$r'$ & $>26$ \cr
$i'$ & $>26$ \cr
NB921 & $\leq23.5$ \cr 
$z'-\textrm{NB921}$ & $\geq1.3$ \cr
\enddata
\label{tab:selection}
\end{deluxetable}

\section{Spectroscopic Follow-Up}
\cite{songaila18} spectroscopically followed up 7 of the candidates, and we followed 
up the remaining 7 with DEIMOS on Keck~II during observing runs in 2018 and 2019. 
We refer the reader to A.~Songaila et al.\ (2020, in preparation), where we 
provide a more detailed description of the data and analysis of the spectra. 
Briefly, we used the G830 grating with a $1''$ slit, which results in a resolution of 
83~km~s$^{-1}$ for the $z=6.6$ LAEs, as measured from sky lines. We took three 20 minute 
sub-exposures for each source using $\pm1.5''$ dithering along the slit for optimal sky subtraction. 
Our total exposure times ranged from 1--3 hours, depending on the source. 

\begin{figure*}[ht]
\includegraphics[width=\linewidth]{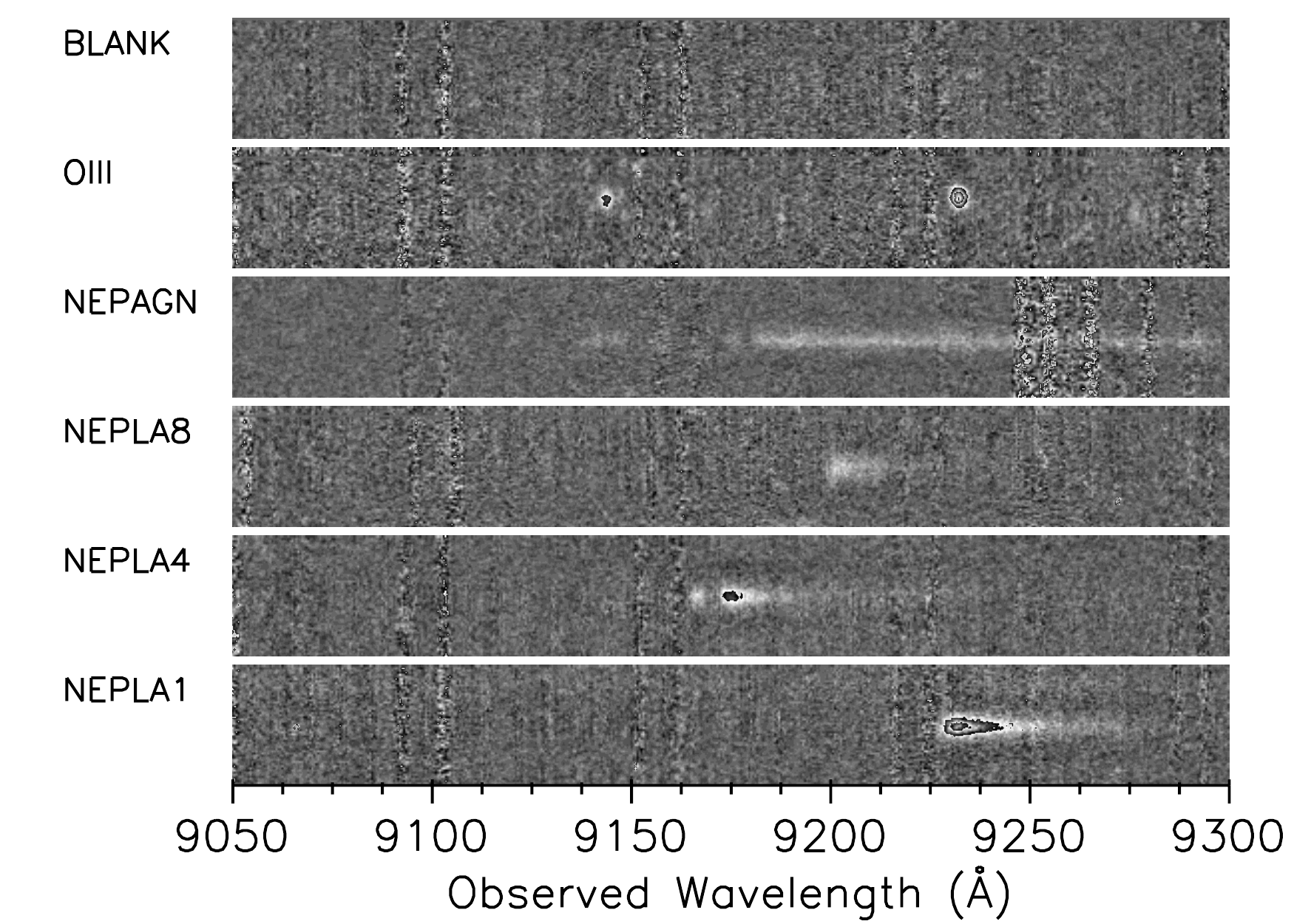}
\caption{2D spectra for six $z=6.6$ LAE candidates with comparable NB921 magnitudes. 
The top spectrum has no detected lines and is one of the two spectroscopic non-detections. 
The second spectrum is one of the two [OIII] emitters, with the [OIII] doublet and fainter
H$\beta$ line visible. The third spectrum 
is the single AGN, featuring a bright continuum with no \Lya line emission. The fourth spectrum is an 
example of a confirmed ULLAE spectrum (NEPLA8) with strong \Lya emission featuring a sharp 
blue break and a red wing. It is one of the least luminous ULLAEs in the sample.
The fifth spectrum is the double-peaked ULLAE NEPLA4. 
The bottom spectrum is the most luminous confirmed ULLAE so far, NEPLA1.}
\label{fig:2dspectra}
\end{figure*}

We carried out the data reduction using the standard pipeline presented in \cite{cowie96}.
In short, we performed an initial sky subtraction, pixel by pixel, using the minimum counts 
recorded by a pixel across the three dithered exposures. We then median added the three frames.  
We rejected cosmic rays using a $3\times3$ median spatial filtering pass. We removed geometric 
distortion from the spectra by tracing brighter objects in the slit mask. We then
calibrated the wavelength scales against the sky lines in the spectra. 
Finally, we performed another sky subtraction. We show a range of two-dimensional (2D) 
spectral images for our sample in Figure~\ref{fig:2dspectra}, including a
spectroscopic non-detection, 
an [OIII] emitter, the single detected AGN (NEPAGN), the double-peaked ULLAE NELPA4, 
and two other ULLAEs: NEPLA1 (the most luminous ULLAE in the sample) 
and NEPLA8 (the second least luminous ULLAE in the sample).

\begin{figure*}[ht]
\includegraphics[width=0.33\linewidth]{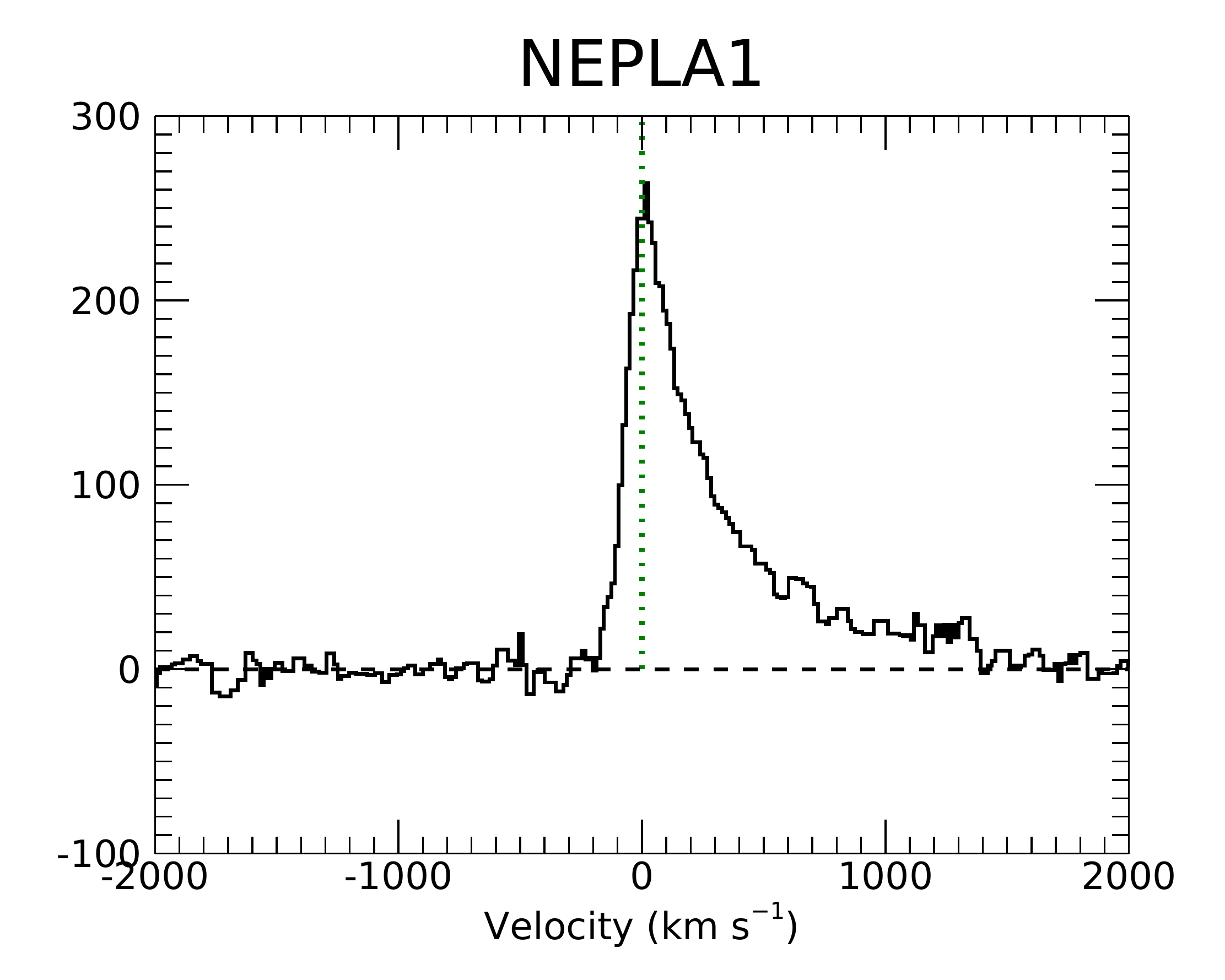}
\includegraphics[width=0.33\linewidth]{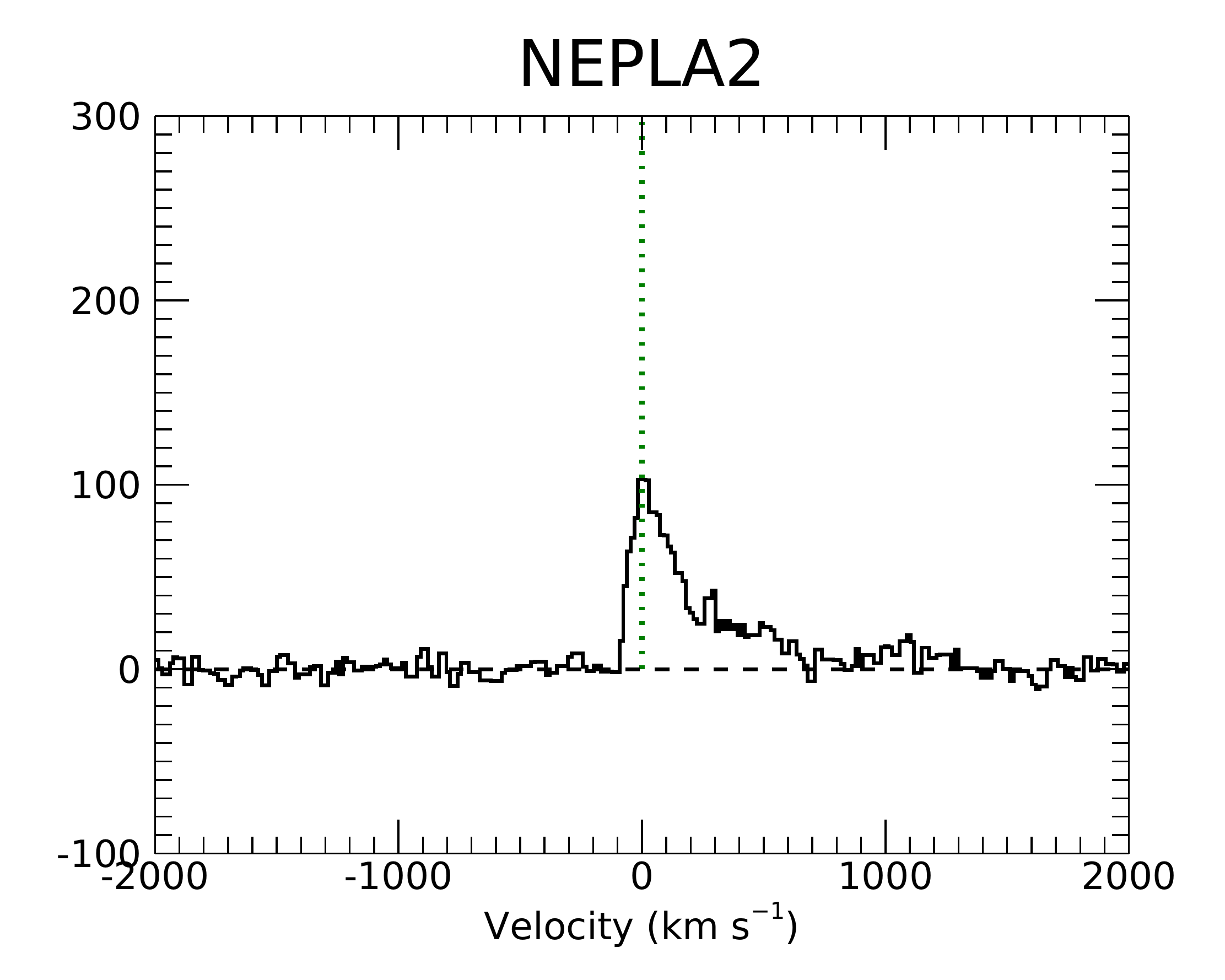}
\includegraphics[width=0.33\linewidth]{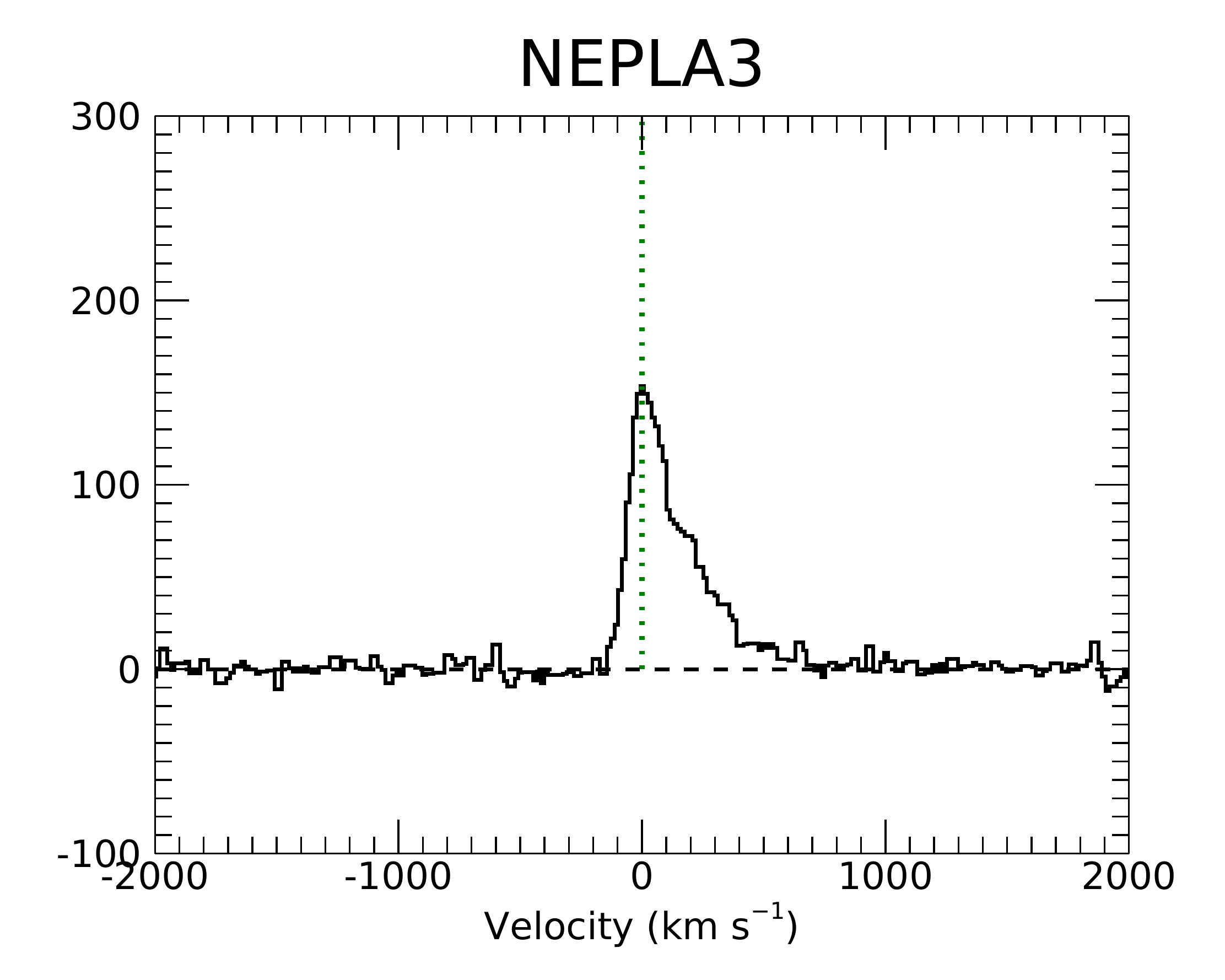}
\includegraphics[width=0.33\linewidth]{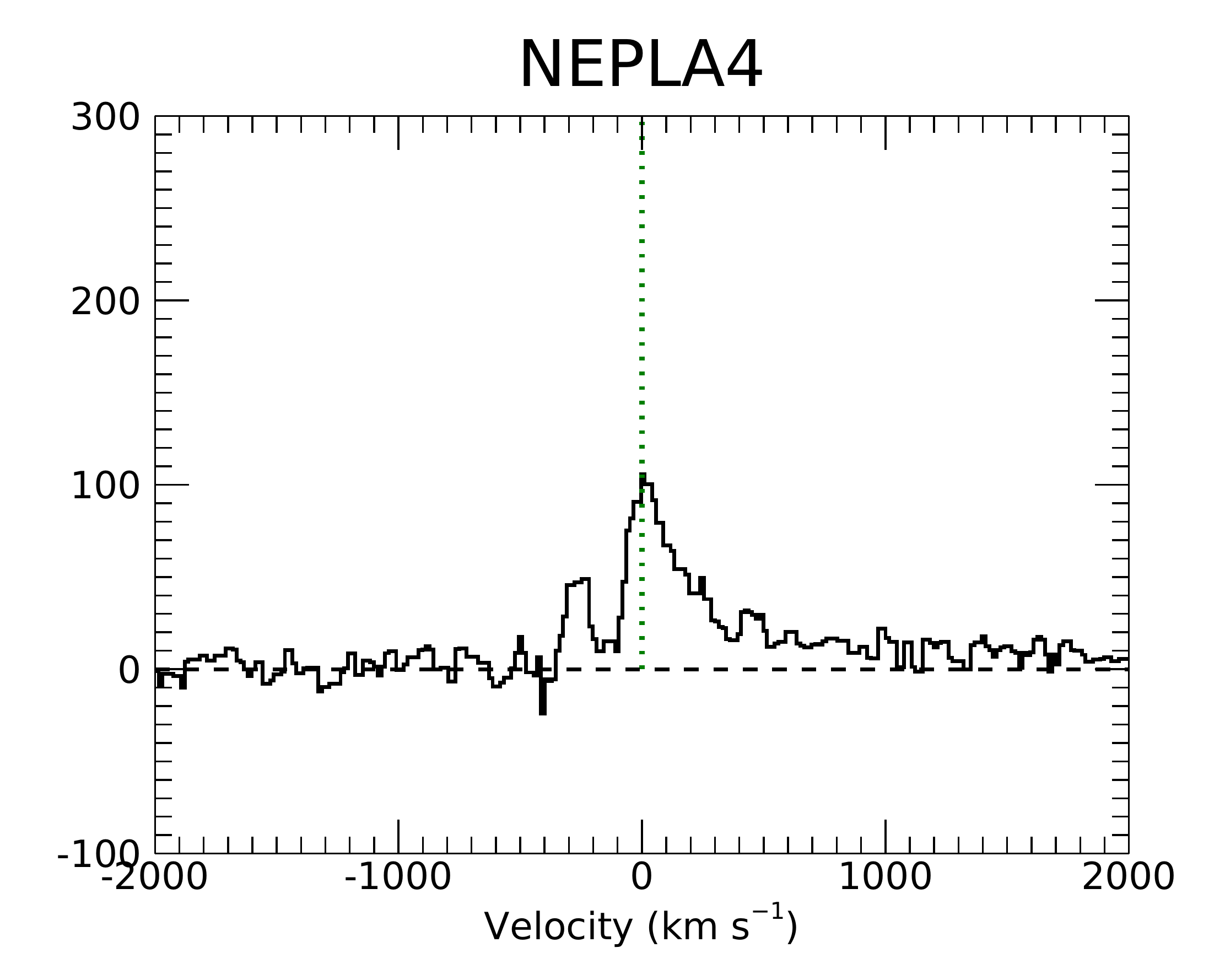}
\includegraphics[width=0.33\linewidth]{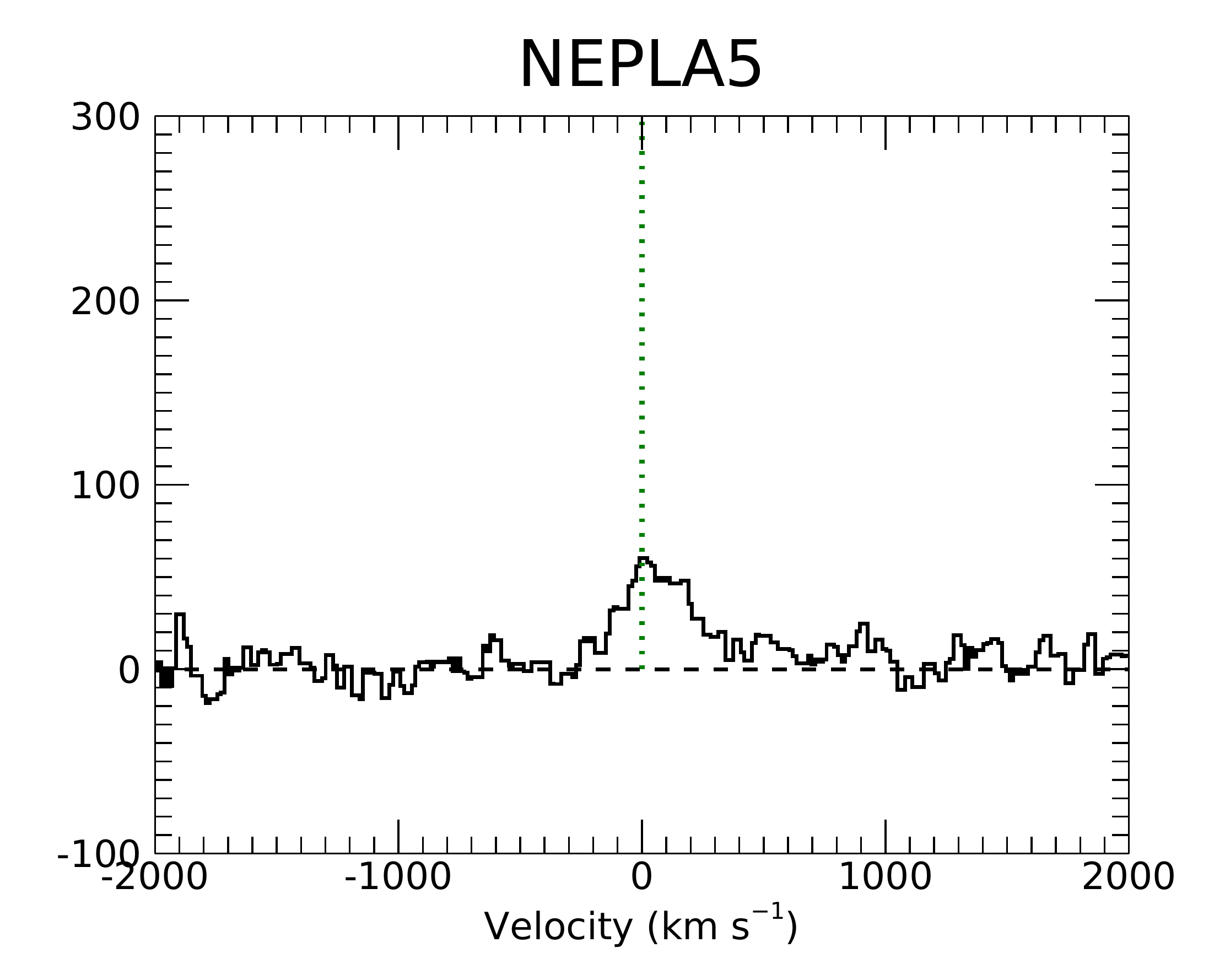}
\includegraphics[width=0.33\linewidth]{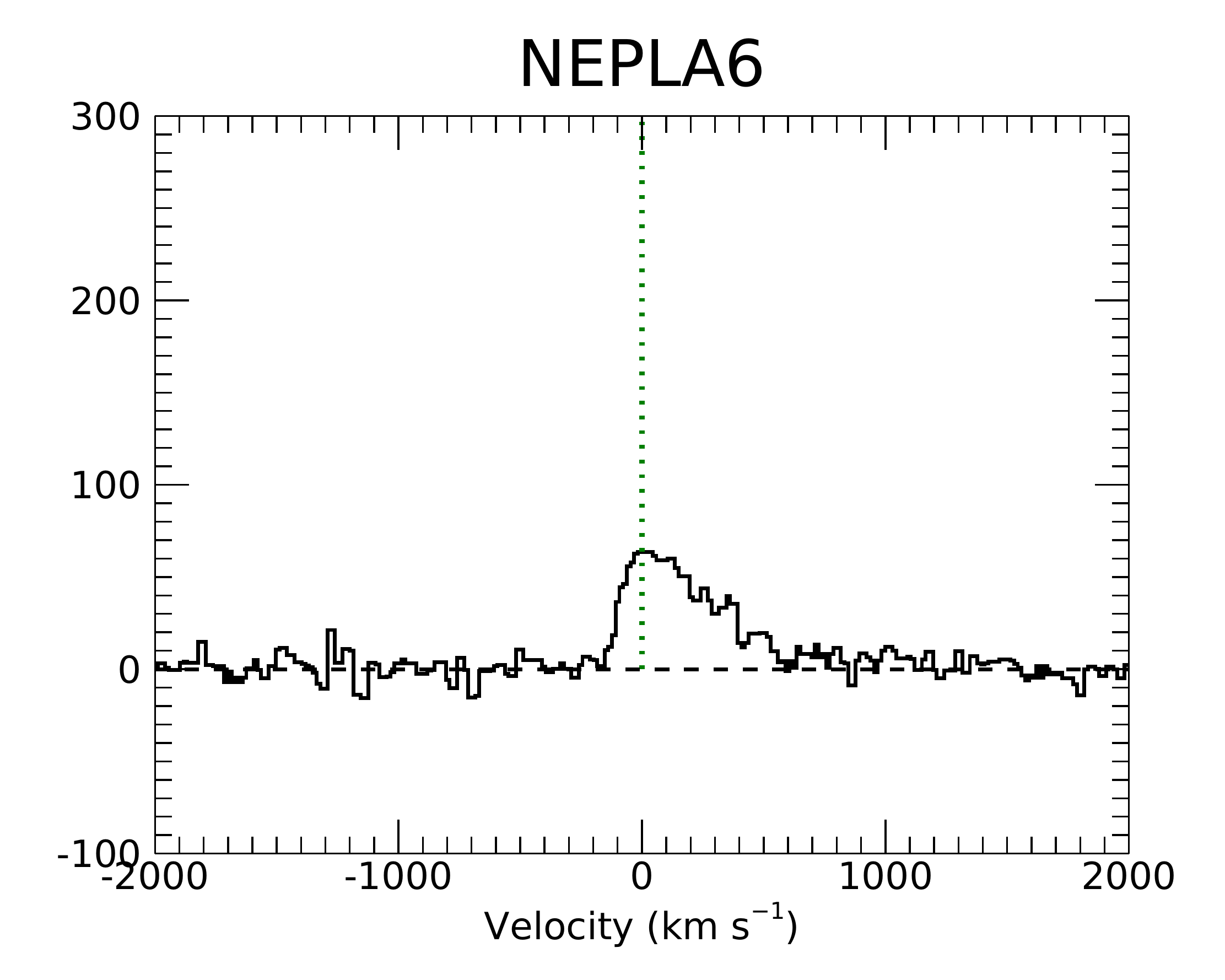}
\includegraphics[width=0.33\linewidth]{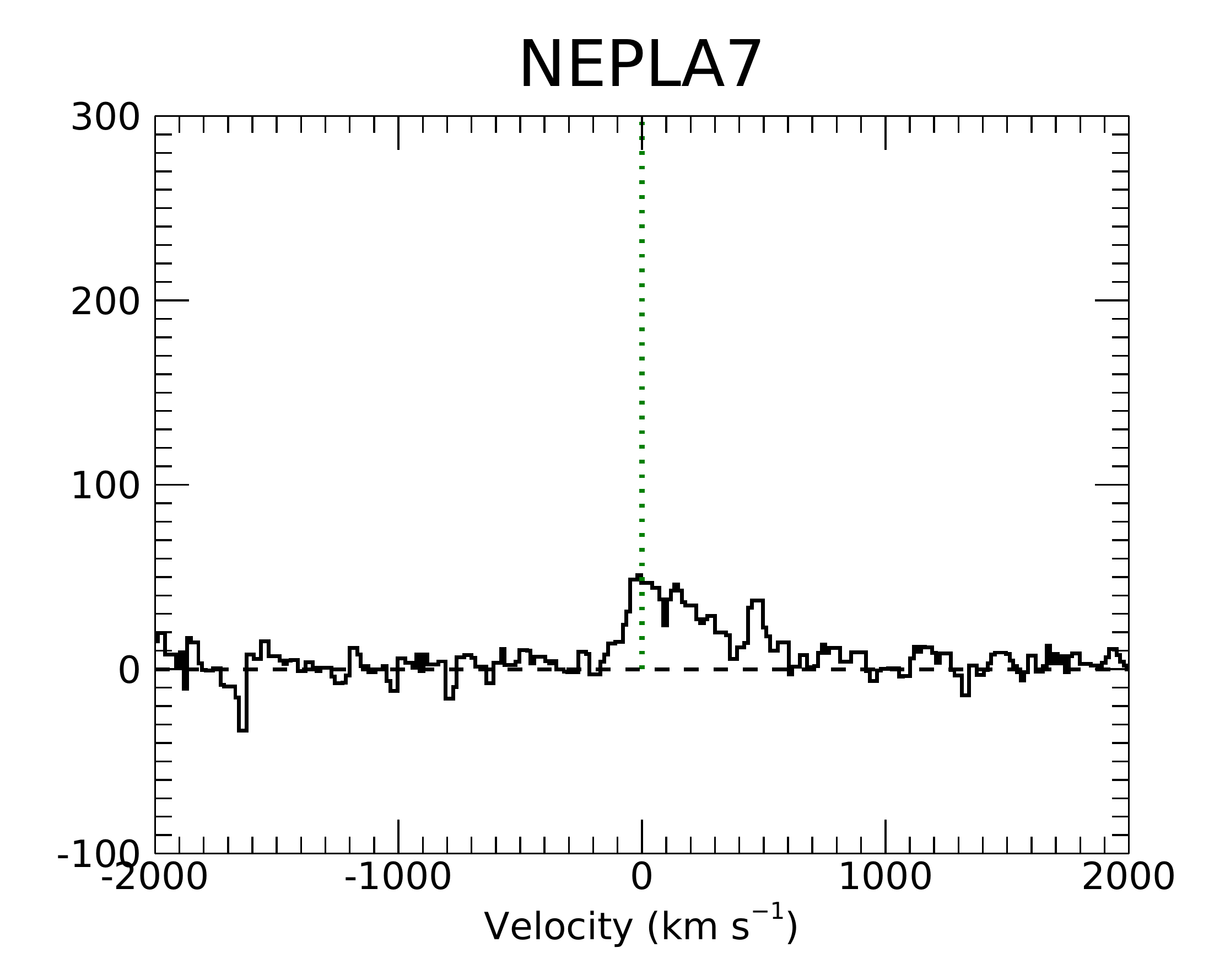}
\includegraphics[width=0.33\linewidth]{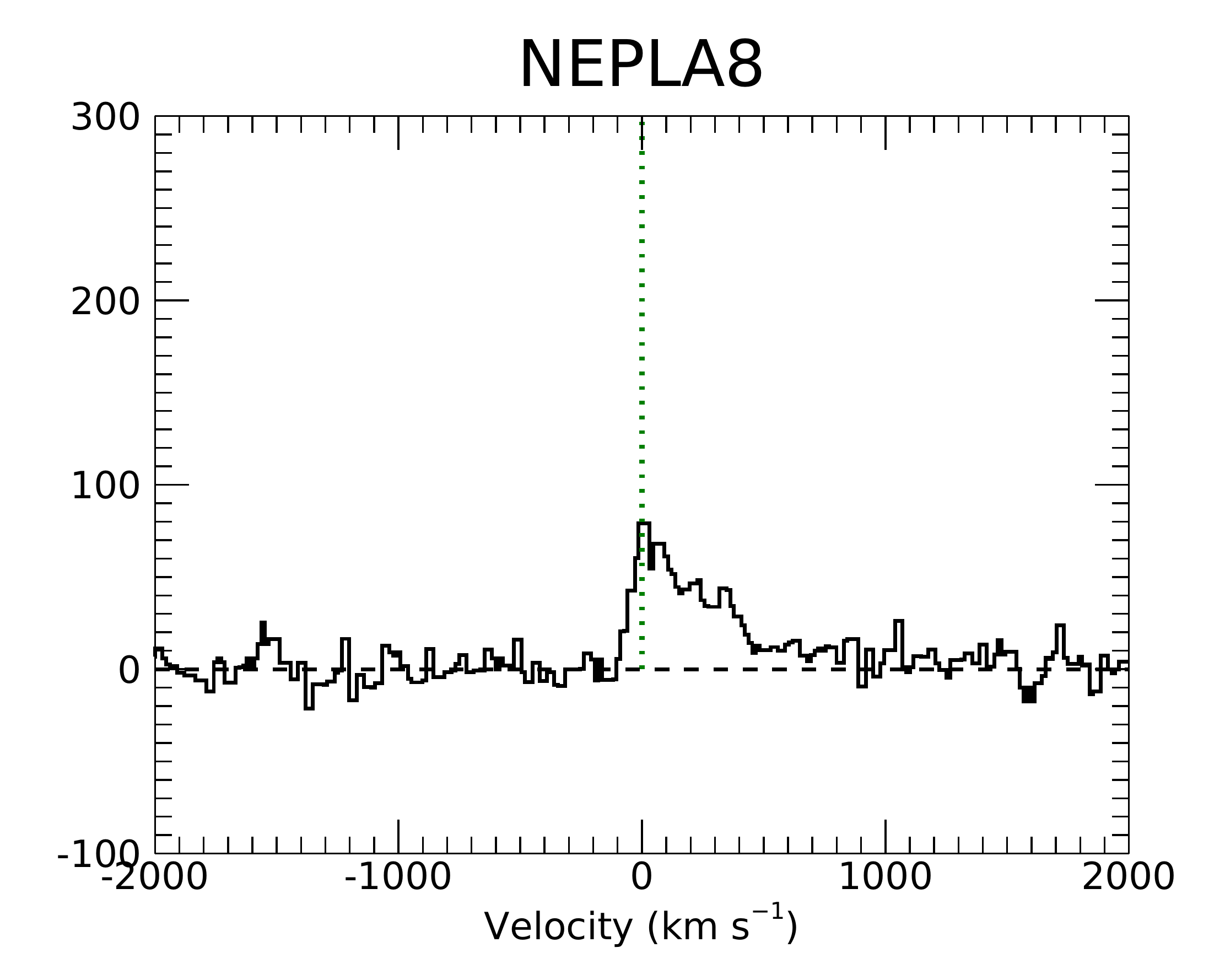}
\includegraphics[width=0.33\linewidth]{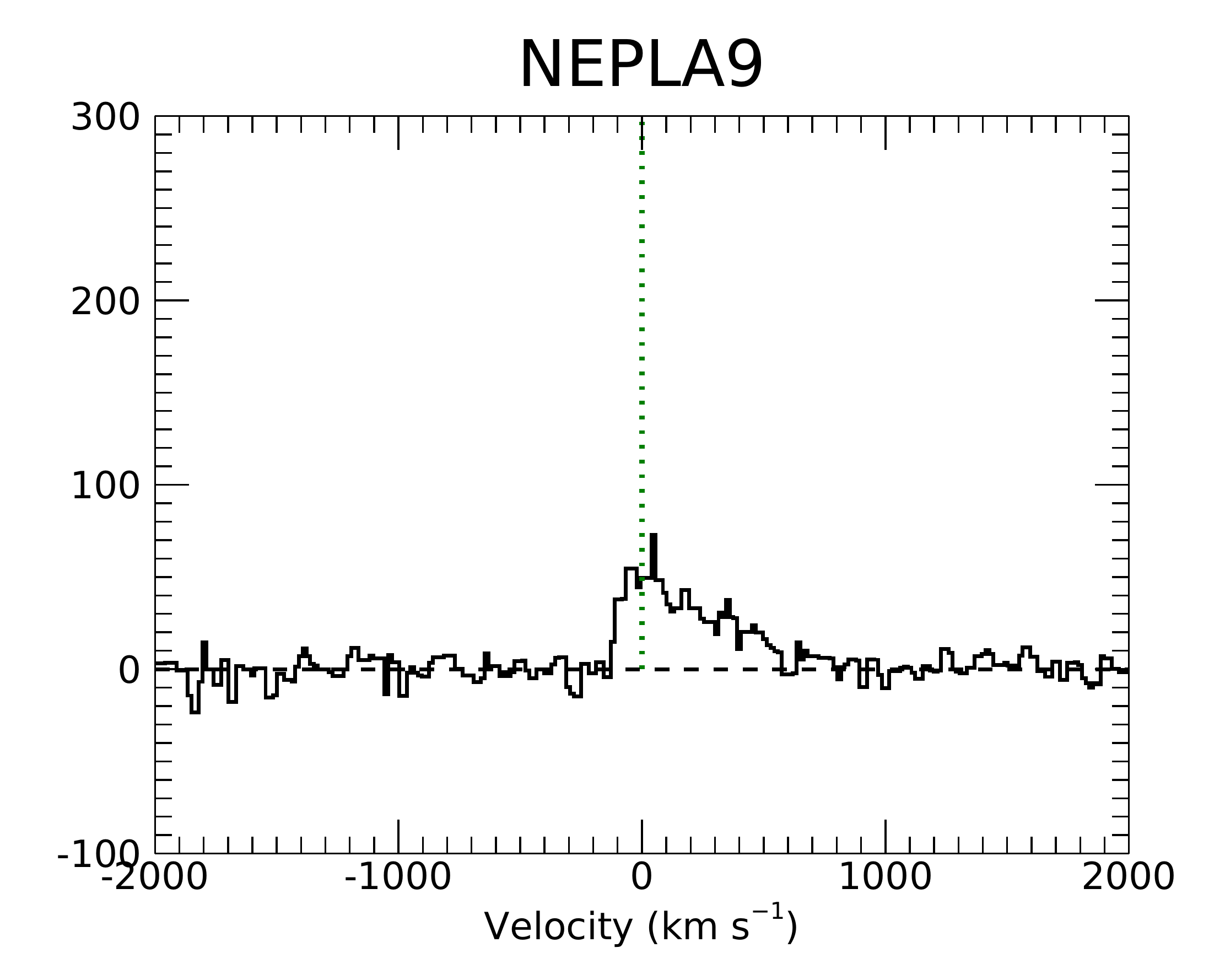}
\caption{1D spectra of the nine confirmed $z=6.6$ ULLAEs in the NEP field. 
The vertical scale is the flux in arbitrary units, but these are
consistent between the spectra, so the normalizations can be
directly compared.
}
\label{fig:1dspectra}
\end{figure*}

In total, we confirmed 9 NEP candidates as $z=6.6$ ULLAEs. 
We present the one-dimensional (1D) spectra of all 9 in Figure~\ref{fig:1dspectra}.
As in \cite{songaila18}, for each profile, we chose the peak value of the \Lya line in the 1D
spectrum as the zero-velocity standard from which to calculate the spectroscopic redshift. 
This is not the galaxy redshift, which may be blueward of this value.  For example,
for the LAE VR7, the \Lya flux peaks at $z=6.534\pm0.001$ \citep{matthee20}, 
which corresponds to a velocity offset of $+213^{+19}_{-20}$~km~s$^{-1}$ relative to the systemic 
redshift traced by [CII]$_{\rm 158~\mu m}$ ($z=6.5285$; \citealt{matthee19}).
The peak redshifts allow us to compare the redshifts of the sample and are adequate for
calculating the transmission throughput needed for determining the line fluxes.

We identified one of the remaining 5 candidates as a $z=6.6$ AGN
\citep[see the 1D spectrum in][]{songaila18}.
This spectrum shows broad \Lya\ emission, along with OVI and NV.
We identified two other candidates as high EW [OIII]$\lambda$5007 emitters at $z=0.84$ and $z=0.85$.
The continuum in these sources is not detected in any of the blue bandpasses, making this type of 
source hard to remove with a purely photometric selection.  
The spectra only show the [OIII] doublet and H$\beta$ line, in each case
with both of the [OIII] lines being much
stronger than the H$\beta$ line (see the second spectrum in Figure~\ref{fig:2dspectra}).
[OIII] emitters are cited as leading sources of 
contamination in photometrically identified LAE samples \citep{konno18}. 
Finally, we did not detect any lines in the last two candidates.

Characterizing the noise in the spectral images is not
straightforward, because of the presence of the residuals
from the sky subtraction. However, the spectra with detected
lines are clearly distinguishable from the two spectra without
(see Figure~\ref{fig:2dspectra}).
In order to quantify this, we used box apertures in the 2D spectra
with 30 spectral pixels
and 10 spatial pixels to measure the distribution of
counts in the wavelength region covered by the NB921 filter
and lying within $3''$ of the nominal position for the NEP candidates.
We then compared the largest value
measured in any of these apertures with the dispersion.
Typically, the S/N of detected \Lya lines is in excess of 10, 
while that of [OIII]$\lambda$5007 lines is greater than 8.
 For the two spectra without detected lines, we did not see
boxes with S/N greater than 2. Thus, these two sources are clearly spurious,
and the normalization of our LF will not be artificially lowered due 
to their removal from the initial sample. 

We note that it is also possible to confirm the reality of 
the LAEs with broadband continuum data in $z'$ and $y'$, but 
the continuum data for the NEP field are not deep enough to do this.

Our follow-up success rate for ULLAEs in the NEP 30~deg$^2$ area is thus 64\%.
This illustrates how critical spectroscopic follow-up is for the confirmation 
of bonafide ULLAEs and the filtration of contaminants.  
We supplement our 9 NEP ULLAEs with two ULLAEs from the COSMOS field:
COLA1 \citep{hu16} and CR7 \citep{sobral15}. COLA1 and CR7 are a spectroscopically complete 
ULLAE sample observed in the same DEIMOS configuration as the NEP sources 
\citep{songaila18}. 

We note that \cite{songaila18} also targeted the LAE MASOSA
\citep{sobral15} in COSMOS. However, its luminosity of 
$\log L(\rm \Lya)=43.42$~erg~s$^{-1}$ lies just below the 
ULLAE luminosity limit, so we do not include it in this study.

\section{Line Fluxes and Luminosities}
\begin{figure}[ht]
\includegraphics[width=\linewidth]{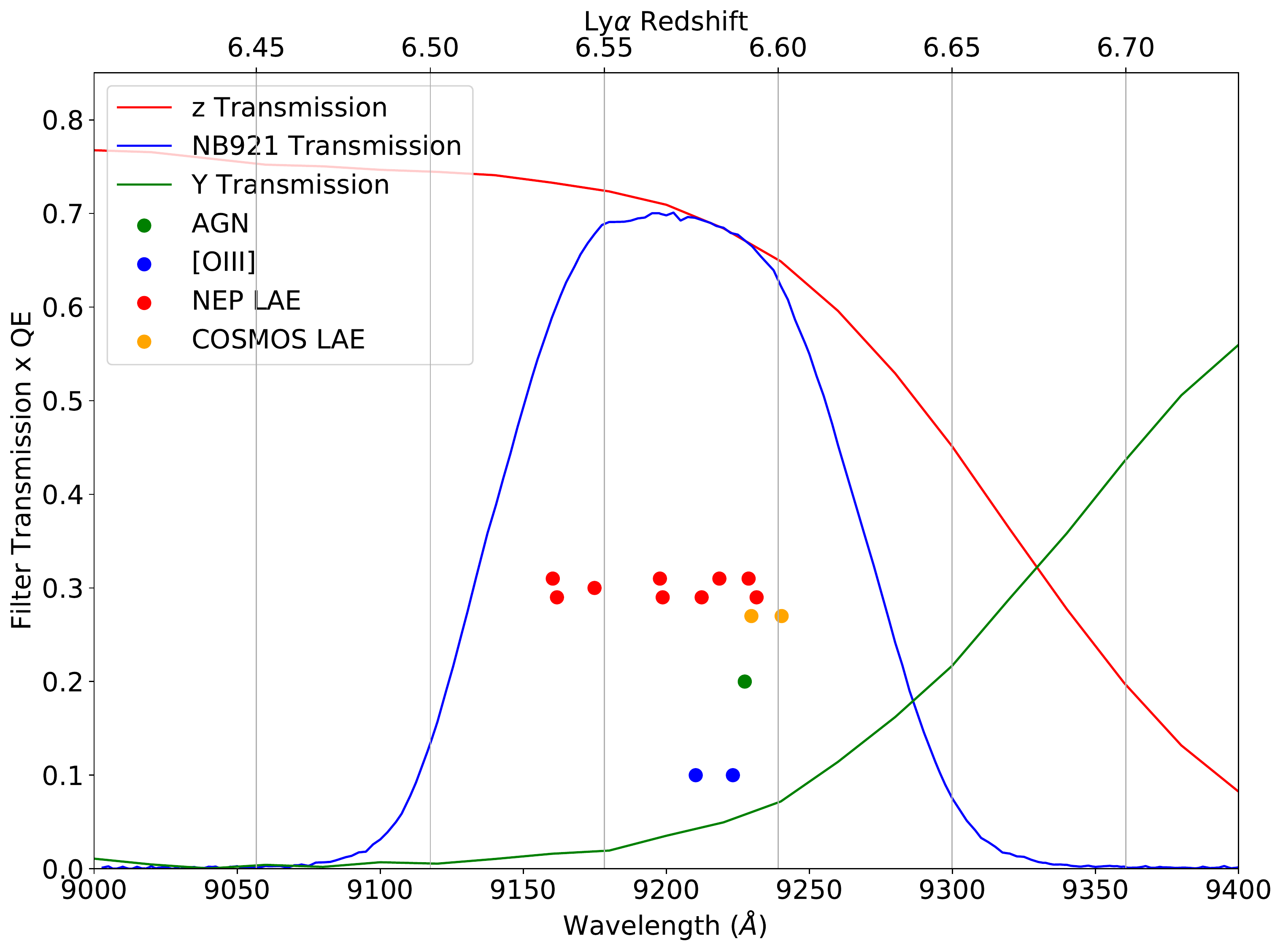}
\caption{Product of the filter transmission and CCD quantum efficiency for the 
HSC filters $z$ (red), NB921 (blue), and $Y$ (green). Red (orange) circles show the 
redshift and observed redshifted peak \Lya wavelength for each NEP (COSMOS) ULLAE. 
Blue circles show the weighted 
mean wavelength of the [OIII] doublet for the two [OIII] emitters. 
The single green circle is a high-redshift AGN. Vertical offsets 
are simply used to distinguish overlapping points.}
\label{fig:filters}
\end{figure}

A primary purpose of this study is to explore the ultraluminous tail of the $z=6.6$ LAE LF. 
As all of our LAEs are spectroscopically confirmed, we have the advantage over other
works of accurate spectroscopic redshifts and \Lya observed wavelengths. 
In Figure~\ref{fig:filters}, we show the redshifts and observed \Lya wavelengths 
for the NEP and COSMOS sources with the HSC filter profiles superimposed.

To calculate the \Lya line flux, we first measured the NB921
magnitude of each LAE using a $2''$ diameter aperture. We then corrected 
this magnitude to a total magnitude using the median offset between $2''$ and $4''$ 
diameter aperture measurements of galaxies in the
22--24 magnitude range in the neighborhood of the source (typically around $-0.2$~mag). 
A.~Songaila et al.\ (2020, in preparation) describe the conversion of these 
NB921 magnitudes to \Lya
line fluxes by assuming that the NB921 flux is produced solely by the \Lya line.
They also find that if one includes the continuum, then it reduces
the \Lya flux and luminosity by a very small amount, 
typically substantially less than 0.1~dex. 
However, we do not have deep enough continuum data in the NEP to do this
accurately.

Finally, we calculated the line luminosity of each LAE using the cosmological 
luminosity distance based on its spectroscopic redshift. 

In Table~\ref{tab:catalog},
we present the final catalog of the 9 NEP LAEs, the 2 COSMOS LAES, the 
NEP AGN, and the 2 [OIII] emitters. All 9 NEP + 2 COSMOS LAEs fit our 
definition of ``ULLAEs" with $\log L$(Ly$\alpha)>43.5$~erg~s$^{-1}$.

\begin{deluxetable*}{cccccc}
\renewcommand\baselinestretch{1.0}
\tablewidth{0pt}
\tablecaption{Properties of the Spectroscopically Observed Sample}
\scriptsize
\tablehead{Source & R.A. & Decl. & NB921 & Redshift & $\log L($\Lya) \\ 
 & (deg) & (deg) & (AB) & & (erg s$^{-1}$) \\
(1) & (2) & (3) & (4) & (5) & (6)}
\startdata
NEPLA1 & 273.73837 & 65.285995 & 22.56 & 6.5938 & 43.92 \cr
NEPLA2 & 263.61490 & 67.593971 & 23.17 & 6.5831 & 43.71 \cr
NEPLA3 & 265.22437 & 65.510361 & 23.19 & 6.5915 & 43.66 \cr
NEPLA4 & 268.29211 & 65.109581 & 22.97 & 6.5472 & 43.76 \cr
NEPLA5 & 269.68964 & 65.944748 & 23.33 & 6.5364 & 43.60 \cr
NEPLA6 & 262.44296 & 65.180443 & 22.92 & 6.5660 & 43.75 \cr
NEPLA7 & 272.66104 & 67.386055 & 23.34 & 6.5780 & 43.59 \cr
NEPLA8 & 262.30838 & 65.599663 & 23.28 & 6.5668 & 43.61 \cr
NEPLA9 & 276.23441 & 67.606667 & 23.39 & 6.5352 & 43.63 \cr
\hline
COLA1 & 150.64751 & 2.2037499 & 23.11 & 6.5923 & 43.70 \cr
CR7 & 150.24167 & 1.8042222 & 23.26 & 6.6010 & 43.67 \cr 
\hline
NEPAGN & 263.17966 & 65.520416 & 22.70 & 6.5904 &  \nodata \cr 
\hline
OIII & 275.35461 & 64.775719 & 23.24 & 0.8465 & \nodata \cr
OIII & 277.74066 & 68.367943 & 22.88 & 0.8439 & \nodata \cr 
\enddata
\label{tab:catalog}
\tablecomments{Columns: (1) Source name, 
(2) and (3) R.A. and decl.,
(4) NB921 $2''$ diameter aperture magnitude corrected to total
\citep[for the COSMOS field, we measured the magnitudes from the 
public HSC Subaru Strategic Program (SSP) NB921 image;][]{aihara19},
(5) redshift, and (6) logarithm of the \Lya line luminosity.}
\end{deluxetable*}

\section{Incompleteness Measurement}
To characterize the incompleteness of our sample, we developed a simulation 
program in which artificial LAEs are inserted into all 7 survey bands. We generate
these artificial LAEs assuming a rest-frame \Lya FWHM line width of 
4~\AA\ with a tunable line luminosity and a flat continuum redward of the \Lya 
line at an intensity of 2.5\% of the \Lya line peak. This spectrum is then redshifted 
to a tunable redshift and is ``observed" by HSC using the filter transmission 
curves and CCD quantum efficiency 
to derive observed magnitudes for the artificial source in each of the 7 bands. 
We use these magnitudes in conjunction with the zeropoint and exposure 
time from the stacked survey images to simulate the total counts expected from
the source were it originally observed in a given survey image. From the total counts,
a 2D Gaussian is constructed, given a FWHM seeing for the image, to produce the 
artificial source. 

We generated 1000 random positions in pixel space for each frame in the survey. 
At each of these positions, we placed an artificial source of a predetermined line 
luminosity and redshift. After inserting the artificial sources into the images, we ran
SExtractor \citep{sextractor}, detecting sources in the NB921 filter 
image and measuring magnitudes for all 7 bands at the NB921 detection positions. 
We then filtered the 
resulting catalog using our magnitude cuts from the original source selection 
($z'$--NB$921\geq1.3$, NB$921<23.5$, $(g', r', i')>26$). We searched this new 
catalog for the known artificial objects, and the percentage of recovered objects 
is then the completeness. We ran this process across all 197 survey images within 
the selected 30~deg$^2$ for all permutations of $z=6.51$--6.63 in intervals of 0.01, 
and $\log L(\rm \Lya)=43.5$--44.0~erg~s$^{-1}$ in intervals of 0.05, simulating 
a total of $>27$ million sources.

\begin{figure}[h]
\includegraphics[width=\linewidth]{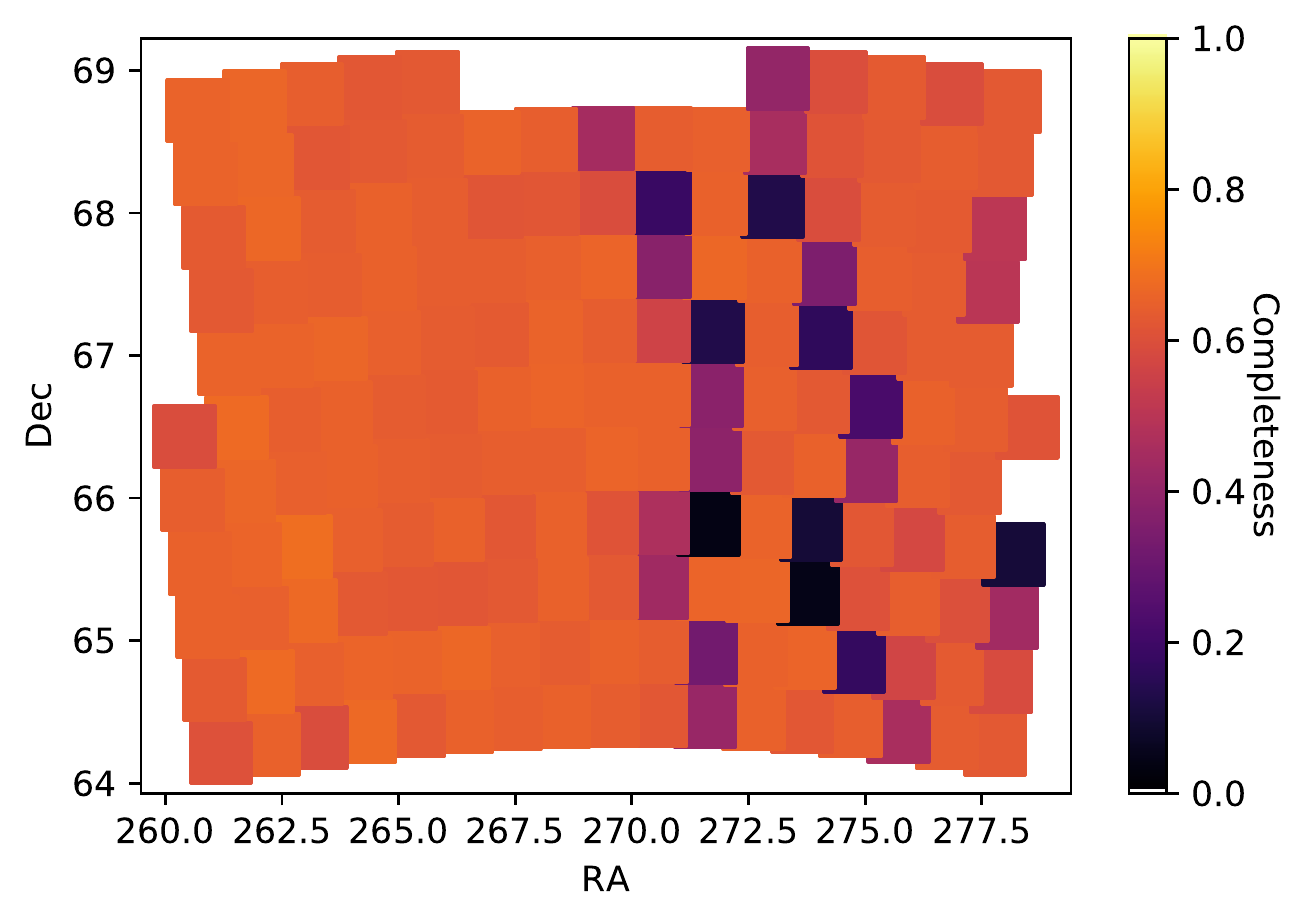}
\caption{Completeness measures averaged over $\log L(\rm \Lya$)=43.5--44.0 and $z=6.52$--6.62 for each skycell in HEROES.}
\label{fig:incskycells}
\end{figure}

\begin{figure}[h]
\includegraphics[width=\linewidth]{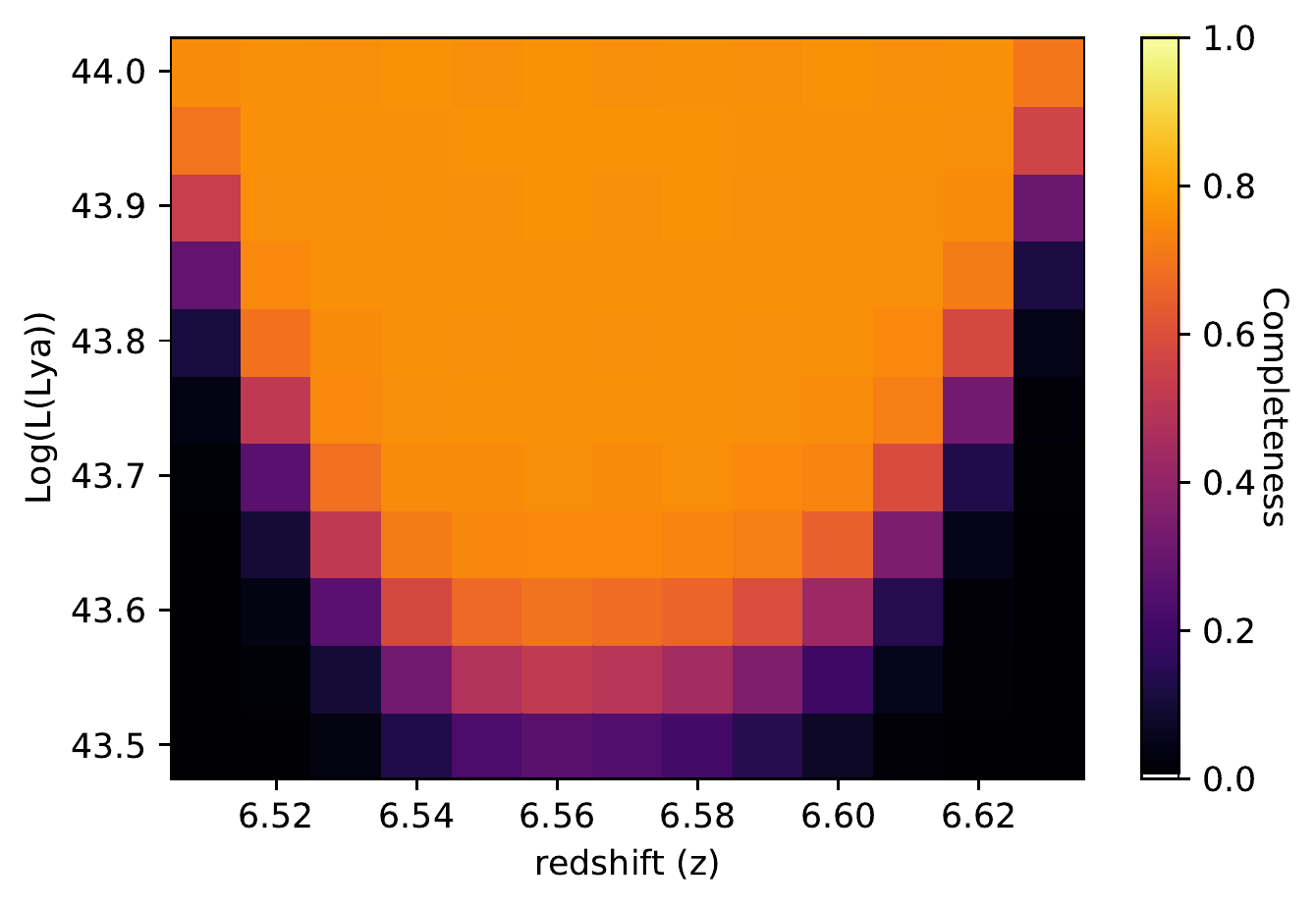}
\caption{Completeness measures averaged over all skycells in HEROES.}
\label{fig:incllyaz}
\end{figure}

We show the results of our incompleteness analysis in Figures~\ref{fig:incskycells} and \ref{fig:incllyaz}. In Figure~\ref{fig:incskycells}, we show the completeness of each skycell 
in the survey averaged over the redshift range $z=6.52$--6.62 and the luminosity 
range $\log L(\rm \Lya)=43.5$-44.0~erg~s$^{-1}$. 
Most skycells have an average total completeness 
of $\sim$63\%. However, some cells along two tracks through the field have sections of missing coverage and thus far lower completeness (a median of 37\%). 
We still consider these cells in our survey, we just use the incompleteness 
measurement to account for the missing area. 

In Figure \ref{fig:incllyaz}, we plot the completeness as a function of redshift and 
\Lya\ luminosity averaging over all skycells. As expected, the filter profile of the 
NB921 filter is visible in the plot as a result of our magnitude cut at NB921$\leq23.5$. 
Thus, insufficiently luminous LAEs at redshifts away from the center of the bandpass 
fail to pass the magnitude cuts, resulting in nearly 0\% completeness for combinations 
of luminosity and redshift significantly far away from the filter transmission peak. 
However, when the luminosity of LAEs is bright enough to pass the magnitude 
selection cuts at a given redshift, the average completeness is nearly 76\%. 
The main sources of incompleteness in this non-magnitude limited regime appear 
to be confusion from proximity to foreground stars, data artifacts, or overlap 
with other field sources. 

We adopt our NEP incompleteness analysis for the COSMOS field.  
As the COSMOS sample of 2 ULLAEs represents only $\sim$18\% of the full 
sample and $\sim$13\% of the surveyed volume, small differences in the 
incompleteness of the two fields would have little effect on the LF.

\section{Luminosity Functions}
We calculate the comoving survey volume from the FWHM bounds of the NB921 
filter in redshift space ($z=6.52$--6.62) and from the 30~deg$^2$ NEP survey area. 
We find a comoving volume of $2.46\times 10^7$~Mpc$^3$. 
For the 4~deg$^2$ COSMOS field 
\citep{hu16}, we find a comoving volume of $3.28\times 10^6$~Mpc$^3$. 
The combined comoving volume is then $2.78\times 10^7$~Mpc$^3$.

We use this volume to construct the $z=6.6$ ULLAE LF at 
$\log L($\Lya)$>43.5$~erg~s$^{-1}$.
Since all of our sources are spectroscopically observed and
have reliable spectroscopic redshifts, the filter transmission profile 
corrections are exact in the determinations of the source luminosities. 
Additionally, we do not need to correct for contaminating sources, such as 
lower-redshift [OIII] emitters or high-redshift AGNs, as would be necessary
in the analysis of purely photometric samples. 
We applied our incompleteness correction by dividing the uncorrected $\log \phi$
of each of the two luminosity bins ($\log L(\textrm{\Lya})=43.5-43.75$
and $43.75-44\textrm{ erg s}\per$) by the overall completeness in each bin.
In Table~\ref{tab:lf}, we give our incompleteness corrected and uncorrected LF measurements 
for the two $\Delta \log L (\textrm{\Lya})=0.25$ bins, as well as the completeness of each bin. 
Our uncertainties are purely Poissonian.

\begin{deluxetable*}{cccc}[ht]
\renewcommand\baselinestretch{1.0}
\tablewidth{0pt}
\tablecaption{Luminosity Function Data}
\tablehead{$\log L(\textrm{\Lya})$ & Uncorrected $\log \phi$ & Corrected $\log \phi$ & Completeness  \cr
& [$\Delta \log L(\textrm{\Lya})]^{-1}$ Mpc$^{-3}$ & [$\Delta \log L(\textrm{\Lya})]^{-1}$ Mpc$^{-3}$ & 
}
\startdata
$43.625$ & $-5.940^{+0.131}_{-0.198}$ & $-5.526^{+0.131}_{-0.198}$ & 0.385  \cr
$43.875$ & $-6.365^{+0.189}_{-0.374}$ & $-6.198^{+0.189}_{-0.374}$ & 0.680  \cr
\enddata
\tablecomments{LF measurements
for bin widths of $\Delta \log L (\textrm{\Lya})=0.25$. 
The uncertainties are Poissonian.}
\label{tab:lf}
\end{deluxetable*}

\begin{figure*}[ht]
\includegraphics[width=\linewidth]{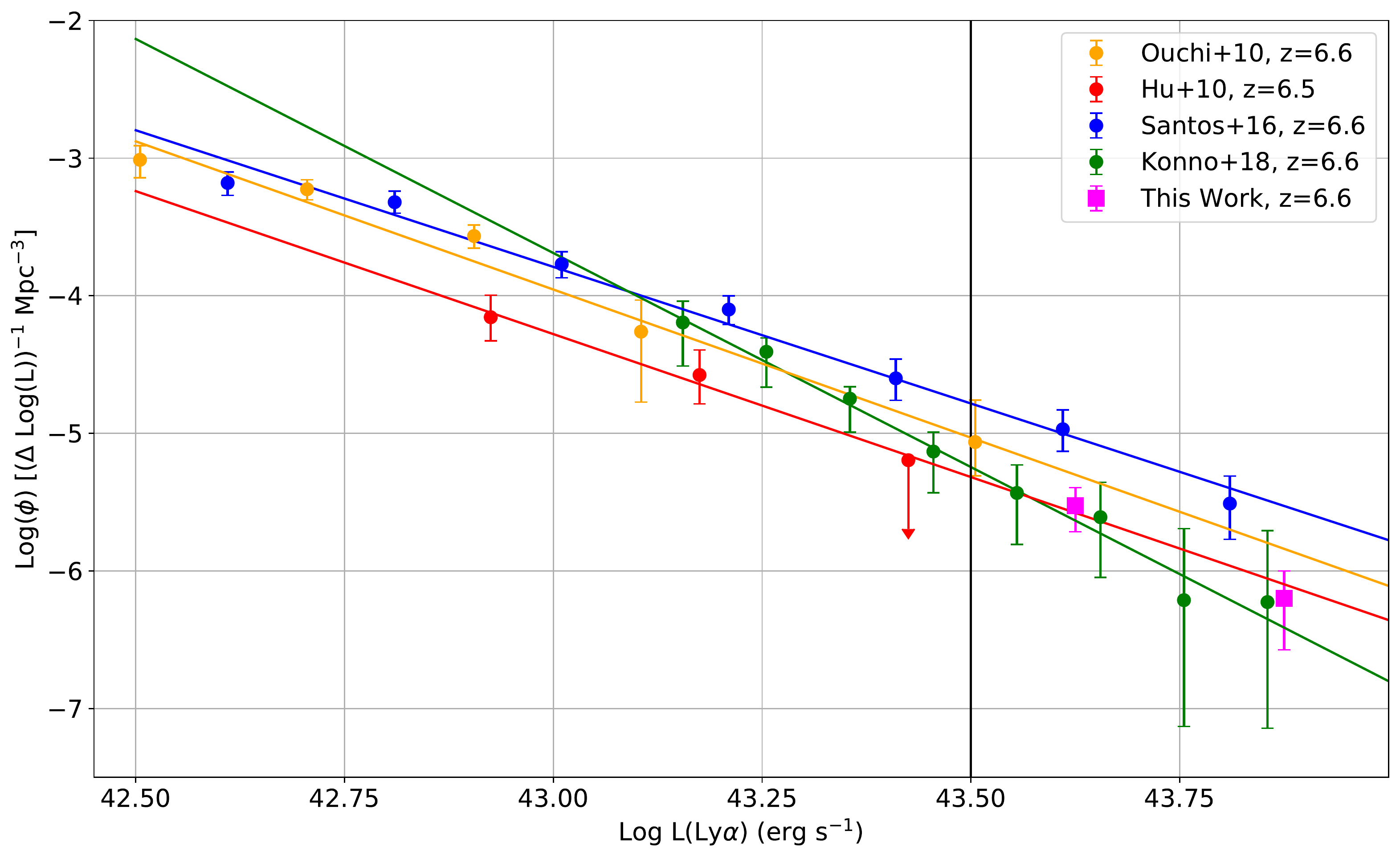}
\includegraphics[width=\linewidth]{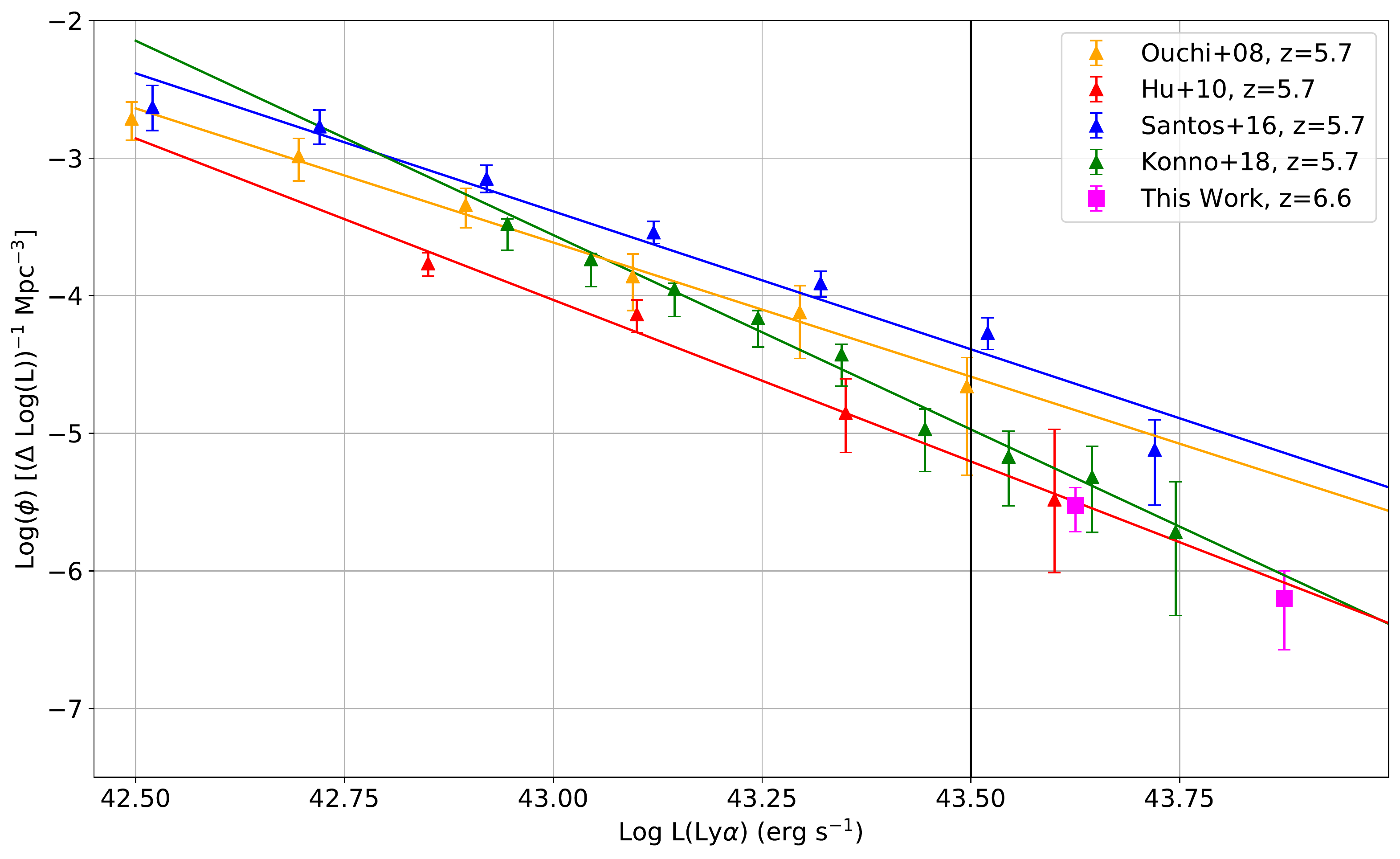}
\caption{
LF measurements for our $z=6.6$ ULLAE sample,
divided into two bins (pink squares): 
$43.5~{\rm erg~s}^{-1}<\log L(\rm \Lya)<43.75$~erg~s$^{-1}$ 
and $43.75~{\rm erg~s}^{-1}<\log L(\rm \Lya)<44.0$~erg~s$^{-1}$.
The vertical line defines our ultraluminous cutoff at 
$\log L(\rm \Lya)> 43.5$~erg~s$^{-1}$.
For comparison, we plot literature LAE LFs at
(a) $z=6.6$ \citep{ouchi10,hu10,santos16,konno18} and
(b) $z=5.7$ \citep{ouchi08,hu10,santos16,konno18}
(see legends for colors).
Note that the $z=6.6$ LF of \cite{santos16} is an updated version
from \cite{matthee15}.
All the literature samples at both redshifts are well represented 
by individual power law fits (matching colored lines).
}
\label{fig:66LF}
\end{figure*}

In Figure~\ref{fig:66LF}, we show these measurements as pink squares, 
along with various (a) $z=6.6$ and (b) $z=5.7$ LAE LFs from the literature,
for comparison (see figure legend for colors). 
For each individual literature sample, we overlay power law fits using free indices 
and normalizations (matching colored lines).
The LF from \cite{konno18} differs strikingly in shape from the remainder, perhaps indicating a problem with the methodology of using a primarily photometric sample from the HSC SSP.  
Excluding \cite{konno18}, the remaining analyses are in fair agreement over the power law index.

Although it is common in the literature to fit Schechter functions, it is readily
apparent from Figure~\ref{fig:66LF} that the power law fits are a good representation 
of the data. The lack of an abrupt fall-off suggests we have not 
yet reached $L_\ast$. Since single power law fits cannot go on indefinitely, 
ultimately the LFs will have to turn down.
This may occur at luminosities just slightly higher than the present ones.

The main point to take away from Figure~\ref{fig:66LF} is that none of the LFs 
from the literature are fully consistent in both power law index and 
normalization with any of the others. 
Thus, for a proper evaluation of whether there is evolution of the LF with redshift,
it is clear that the samples 
at the two redshifts need to be collected and analyzed in exactly the same way. 

In our case, this means we can compare with the 
spectroscopically confirmed $z=6.6$ and $z=5.7$ 
LAE LFs of \cite{hu10}, though they only have
one ultraluminous point at $z=5.7$, and it has large uncertainties.
Indeed, our $z=6.6$ ULLAE LF matches very well to 
both of their LFs. 
This would suggest that there is no evolution in the LFs at the 
ultraluminous end over this redshift range. However, we need more 
spectroscopically confirmed ULLAEs 
at $z=5.7$ to make a definitive statement.

For ease of comparison, we show in Figure~\ref{fig:66LF} the various LFs 
for a fixed power law index of $-2$ and logarithmic normalization of 82.0544 in units
of $[\Delta \log L(\textrm{\Lya})]^{-1}$ Mpc$^{-3}$, taken from 
the best fit to \cite{ouchi10} assuming a fixed power law index of $-2$. 
This time we also separate them into four panels, each of which contains the $z=5.7$ and $z=6.6$
LAE LFs from a single group, along with our $z=6.6$ ULLAE LF (pink squares) for comparison. 

In the first panel, we show the LFs of \cite{ouchi08} and \cite{ouchi10}.
They do not have any data at the ultraluminous end, so we
cannot make a direct comparison with our ULLAE LF. 
They claim a decrease from $z=5.7$ to $z=6.6$ at the $>90$\% 
confidence level at the lower luminosity end, with the $z=6.6$
luminosity density about 30\% of the $z=5.7$ luminosity density.

In the second panel, we show the LFs of \cite{hu10} where, as we noted
above, we are in agreement with their single $z=5.7$ point at the
ultraluminous end (suggesting no evolution), though their point has large 
uncertainties. In their LF comparison, they found a multiplicative factor of 
two decrease in the number density of LAEs from $z=5.7$ to $z=6.5$.

In the third panel, we show the LFs of \cite{santos16}.  They have several 
data points at both redshifts at the ultraluminous end, but their
LFs are 0.3--0.6 dex higher than ours. \cite{konno18} also noted that the
\cite{santos16} number densities of LAEs at all luminosities were too high.
\cite{santos16} (and previously \citealt{matthee15} at $z=6.6$) 
studied the COSMOS, UDS, 
and SA22 fields with Suprime-Cam using a primarily photometric sample. 
Thus, the differences in normalization could be due to their assumptions
about the redshifts in making the filter transmission profile corrections, flux systematics, 
and/or the presence of contaminants that are inherently absent in our 
spectroscopically confirmed sample. 
They find no evolution in the number density at the ultraluminous end
from $z=5.7$ to $z=6.6$, but they see a significant decline at the lower luminosity 
end (by $0.5\pm 0.1$~dex). This drop is similar to that seen by \cite{ouchi08}
and \cite{ouchi10} and by \cite{hu10}.

Finally, in the fourth panel, we show the LFs of \cite{konno18}. 
Their shape inconsistencies with the other surveys are strongly apparent 
in this figure. They have multiple data points at both redshifts at the 
ultraluminous end (though with significant uncertainties), and
we find good agreement with their $z=6.6$ measurements.  Again,
they see evidence for a decrease at the lower luminosity end from $z=5.7$ to 
$z=6.6$ at the $>90$\% confidence level. The evolution they derive is similar to
that reported by the other groups. They make no claim about the evolution
of the LFs at the ultraluminous end.

As discussed above,
given the discrepancies between the different groups' LFs, one needs to exercise
caution in making claims about the 
evolution of the LAE LF from $z=5.7$ to $z=6.6$ unless one is comparing  
samples taken and analyzed in exactly the same way. For us, this means we
can make comparisons with the \citet{hu10} samples at both redshifts; however, 
they are primarily at non-ULLAE luminosities.
In future work, we plan to follow up spectroscopically a population of $z=5.7$ 
ULLAE candidates selected from HEROES in order to construct the LF at the
ultraluminous end.
Then we will be able to make direct comparisons with the 
spectroscopically confirmed $z=6.6$ ULLAE sample presented in this paper,
and hence determine with our own data set whether there is any evolution from 
$z=5.7$ to $z=6.6$ at the ultraluminous end.

\begin{figure*}[htb]
\includegraphics[angle=0,width=\linewidth]{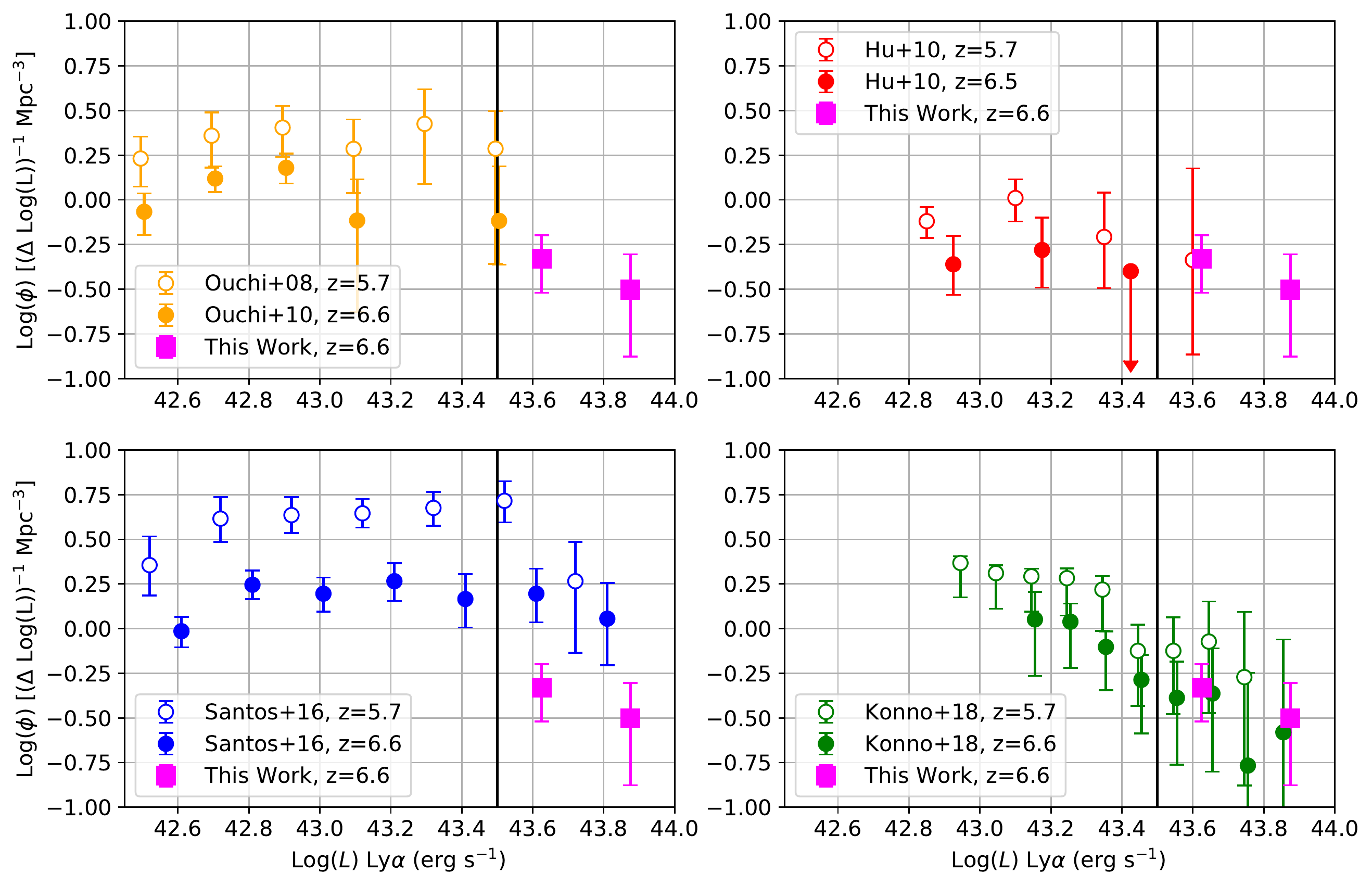}
\caption{Evolution of the \Lya LF from $z=5.7$ to $z=6.6$ for four different
surveys, all shown relative to our $z=6.6$ ULLAE LF (pink squares).
Minute shifts of $\pm 0.02$ in $\log L(\rm \Lya)$ have been made to avoid 
overlap of the data points.}
\label{fig:5766}
\end{figure*}

\section{Summary}

The primary results from our work are as follows:
\begin{itemize}

\item{Using 30~deg$^2$ of deep Subaru HSC $g'$, $r'$, $i'$, $z'$, and NB921 imaging of 
the NEP field, we identified 14 $z=6.6$ ULLAE candidates.}

\item{We spectroscopically observed with Keck DEIMOS 7 of the candidates, and the remaining 
7 were previously observed by \cite{songaila18}.  This provides 9 apectroscopically 
confirmed $z=6.6$ ULLAEs in the NEP field.}

\item{We supplemented the 9 NEP ULLAEs with two spectroscopically confirmed 
$z=6.6$ ULLAEs in the COSMOS field for a total sample of 11. After applying corrections 
for the narrowband filter transmission profile and incompleteness, 
we constructed the $z=6.6$ ULLAE LF from this sample.}

\item{We compared our $z=6.6$ ULLAE LF with $z=5.7$ and $z=6.6$ LAE LFs from the literature.
We showed that none of the literature LFs are fully consistent in both power law index and normalization 
with any of the others, with the \cite{santos16} LFs, in particular, being too high, and the \cite{konno18} LFs 
having an odd shape.}

\item{Given the variations in the literature LFs, it is clear that to determine the evolution
of the LAE LF from $z=5.7$ to $z=6.6$, one 
should only compare LFs at the two redshifts that have been constructed in exactly the same way.
We therefore compared our $z=6.6$ ULLAE LF with the $z=5.7$ and $z=6.6$ LAE LFs of \citet{hu10}, 
which are also spectroscopically confirmed. We found that ours matched very well to both of theirs,
suggesting no evolution in the LFs at the ultraluminous end.}

\item{However, \citet{hu10} only have one ultraluminous point at $z=5.7$, and it has large
uncertainties. Thus, we are working on a spectroscopically complete ULLAE LF analysis 
at $z=5.7$ that should allow us to determine within our own data set whether the ULLAE LF 
evolves over this redshift range.}

\end{itemize}

\section{Acknowledgments} 
We thank the anonymous referee for a constructive report that helped us to improve this paper. 
We gratefully acknowledge support for this research from Jeff and Judy Diermeier through 
a Diermeier Fellowship (A.J.T.), NSF grants AST-1716093 (E.M.H., A.S.) and 
AST-1715145 (A.J.B), the trustees of the William F. Vilas Estate (A.J.B.), and the 
University of Wisconsin-Madison, Office of the Vice Chancellor for Research and 
Graduate Education with funding from the Wisconsin Alumni Research Foundation (A.J.B.). 

This paper is based in part on data collected from the Subaru Telescope.
The Hyper Suprime-Cam (HSC) collaboration includes the astronomical communities 
of Japan and Taiwan, and Princeton University. The HSC instrumentation and software 
were developed by the National Astronomical Observatory of Japan (NAOJ), the Kavli 
Institute for the Physics and Mathematics of the Universe (Kavli IPMU), the University 
of Tokyo, the High Energy Accelerator Research Organization (KEK), the Academia 
Sinica Institute for Astronomy and Astrophysics in Taiwan (ASIAA), and Princeton 
University. Funding was contributed by the FIRST program from Japanese Cabinet 
Office, the Ministry of Education, Culture, Sports, Science and Technology (MEXT), 
the Japan Society for the Promotion of Science (JSPS), Japan Science and 
Technology Agency (JST), the Toray Science Foundation, NAOJ, Kavli IPMU, KEK, 
ASIAA, and Princeton University. 
The NB921 filter was supported by KAKENHI (23244025) Grant-in-Aid for Scientific 
Research (A) through the Japan Society for the Promotion of Science (JSPS). 

This paper also makes use of data collected at the Subaru Telescope and retrieved 
from the HSC data archive system, which is operated by Subaru Telescope and 
Astronomy Data Center at National Astronomical Observatory of Japan. 
Data analysis was in part carried out with the cooperation of Center for 
Computational Astrophysics, National Astronomical Observatory of Japan.

This paper is based in part on data collected from the Keck~II Telescope.
The W.~M.~Keck Observatory is operated as a scientific partnership among the 
California Institute of Technology, the University of California, and NASA, and was 
made possible by the generous financial support of the W.~M.~Keck Foundation.

The authors wish to recognize and acknowledge the very significant cultural role and 
reverence that the summit of Maunakea has always had within the indigenous 
Hawaiian community. We are most fortunate to have the opportunity to conduct 
observations from this mountain.

\bibliography{Lyabib.bib}

\begin{thebibliography}{}
\expandafter\ifx\csname natexlab\endcsname\relax\def\natexlab#1{#1}\fi
\providecommand{\url}[1]{\href{#1}{#1}}
\providecommand{\dodoi}[1]{doi:~\href{http://doi.org/#1}{\nolinkurl{#1}}}
\providecommand{\doeprint}[1]{\href{http://ascl.net/#1}{\nolinkurl{http://ascl.net/#1}}}
\providecommand{\doarXiv}[1]{\href{https://arxiv.org/abs/#1}{\nolinkurl{https://arxiv.org/abs/#1}}}

\bibitem[{{Aihara} {et~al.}(2019){Aihara}, {AlSayyad}, {Ando}, {Armstrong},
  {Bosch}, {Egami}, {Furusawa}, {Furusawa}, {Goulding}, {Harikane}, {Hikage},
  {Ho}, {Hsieh}, {Huang}, {Ikeda}, {Imanishi}, {Ito}, {Iwata}, {Jaelani},
  {Kakuma}, {Kawana}, {Kikuta}, {Kobayashi}, {Koike}, {Komiyama}, {Li},
  {Liang}, {Lin}, {Luo}, {Lupton}, {Lust}, {MacArthur}, {Matsuoka}, {Mineo},
  {Miyatake}, {Miyazaki}, {More}, {Murata}, {Namiki}, {Nishizawa}, {Oguri},
  {Okabe}, {Okamoto}, {Okura}, {Ono}, {Onodera}, {Onoue}, {Osato}, {Ouchi},
  {Shibuya}, {Strauss}, {Sugiyama}, {Suto}, {Takada}, {Takagi}, {Takata},
  {Takita}, {Tanaka}, {Terai}, {Toba}, {Uchiyama}, {Utsumi}, {Wang}, {Wang}, \&
  {Yamada}}]{aihara19}
{Aihara}, H., {AlSayyad}, Y., {Ando}, M., {et~al.} 2019, \pasj, 71, 114

\bibitem[{{Bertin} \& {Arnouts}(1996)}]{sextractor}
{Bertin}, E., \& {Arnouts}, S. 1996, \aaps, 117, 393

\bibitem[{{Cowie} {et~al.}(1996){Cowie}, {Songaila}, {Hu}, \&
  {Cohen}}]{cowie96}
{Cowie}, L.~L., {Songaila}, A., {Hu}, E.~M., \& {Cohen}, J.~G. 1996, \aj, 112,
  839

\bibitem[{{Hu} {et~al.}(2010){Hu}, {Cowie}, {Barger}, {Capak}, {Kakazu}, \&
  {Trouille}}]{hu10}
{Hu}, E.~M., {Cowie}, L.~L., {Barger}, A.~J., {et~al.} 2010, \apj, 725, 394

\bibitem[{{Hu} {et~al.}(2004){Hu}, {Cowie}, {Capak}, {McMahon}, {Hayashino}, \&
  {Komiyama}}]{hu04}
{Hu}, E.~M., {Cowie}, L.~L., {Capak}, P., {et~al.} 2004, \aj, 127, 563

\bibitem[{{Hu} {et~al.}(1998){Hu}, {Cowie}, \& {McMahon}}]{hu98}
{Hu}, E.~M., {Cowie}, L.~L., \& {McMahon}, R.~G. 1998, \apjl, 502, L99

\bibitem[{{Hu} {et~al.}(2016){Hu}, {Cowie}, {Songaila}, {Barger},
  {Rosenwasser}, \& {Wold}}]{hu16}
{Hu}, E.~M., {Cowie}, L.~L., {Songaila}, A., {et~al.} 2016, \apjl, 825, L7

\bibitem[{{Hu} {et~al.}(2019){Hu}, {Wang}, {Zheng}, {Malhotra}, {Rhoads},
  {Infante}, {Barrientos}, {Yang}, {Jiang}, {Kang}, {Perez}, {Wold}, {Hibon},
  {Jiang}, {Khostovan}, {Valdes}, {Walker}, {Galaz}, {Coughlin}, {Harish},
  {Kong}, {Pharo}, \& {Zheng}}]{LAGER19}
{Hu}, W., {Wang}, J., {Zheng}, Z.-Y., {et~al.} 2019, \apj, 886, 90

\bibitem[{{Itoh} {et~al.}(2018){Itoh}, {Ouchi}, {Zhang}, {Inoue}, {Mawatari},
  {Shibuya}, {Harikane}, {Ono}, {Kusakabe}, {Shimasaku}, {Fujimoto}, {Iwata},
  {Kajisawa}, {Kashikawa}, {Kawanomoto}, {Komiyama}, {Lee}, {Nagao}, \&
  {Taniguchi}}]{itoh18}
{Itoh}, R., {Ouchi}, M., {Zhang}, H., {et~al.} 2018, \apj, 867, 46

\bibitem[{{Jiang} {et~al.}(2017){Jiang}, {Shen}, {Bian}, {Zheng}, {Wu},
  {Oyarz{\'u}n}, {Blanc}, {Fan}, {Ho}, {Infante}, {Wang}, {Wu}, {Mateo},
  {Bailey}, {Crane}, {Olszewski}, {Shectman}, {Thompson}, \&
  {Walker}}]{jaing17}
{Jiang}, L., {Shen}, Y., {Bian}, F., {et~al.} 2017, \apj, 846, 134

\bibitem[{{Kashikawa} {et~al.}(2011){Kashikawa}, {Shimasaku}, {Matsuda},
  {Egami}, {Jiang}, {Nagao}, {Ouchi}, {Malkan}, {Hattori}, {Ota}, {Taniguchi},
  {Okamura}, {Ly}, {Iye}, {Furusawa}, {Shioya}, {Shibuya}, {Ishizaki}, \&
  {Toshikawa}}]{kashikawa11}
{Kashikawa}, N., {Shimasaku}, K., {Matsuda}, Y., {et~al.} 2011, \apj, 734, 119

\bibitem[{{Konno} {et~al.}(2014){Konno}, {Ouchi}, {Ono}, {Shimasaku},
  {Shibuya}, {Furusawa}, {Nakajima}, {Naito}, {Momose}, {Yuma}, \&
  {Iye}}]{konno14}
{Konno}, A., {Ouchi}, M., {Ono}, Y., {et~al.} 2014, \apj, 797, 16

\bibitem[{{Konno} {et~al.}(2018){Konno}, {Ouchi}, {Shibuya}, {Ono},
  {Shimasaku}, {Taniguchi}, {Nagao}, {Kobayashi}, {Kajisawa}, {Kashikawa},
  {Inoue}, {Oguri}, {Furusawa}, {Goto}, {Harikane}, {Higuchi}, {Komiyama},
  {Kusakabe}, {Miyazaki}, {Nakajima}, \& {Wang}}]{konno18}
{Konno}, A., {Ouchi}, M., {Shibuya}, T., {et~al.} 2018, \pasj, 70, S16

\bibitem[{{Magnier} {et~al.}(2016){Magnier}, {Schlafly}, {Finkbeiner}, {Tonry},
  {Goldman}, {R{\"o}ser}, {Schilbach}, {Chambers}, {Flewelling}, \&
  {Huber}}]{magnier16}
{Magnier}, E.~A., {Schlafly}, E.~F., {Finkbeiner}, D.~P., {et~al.} 2016, arXiv
  e-prints, arXiv:1612.05242

\bibitem[{{Matthee} {et~al.}(2020){Matthee}, {Sobral}, {Gronke}, {Pezzulli},
  {Cantalupo}, {R{\"o}ttgering}, {Darvish}, \& {Santos}}]{matthee20}
{Matthee}, J., {Sobral}, D., {Gronke}, M., {et~al.} 2020, \mnras, 492, 1778

\bibitem[{{Matthee} {et~al.}(2015){Matthee}, {Sobral}, {Santos},
  {R{\"o}ttgering}, {Darvish}, \& {Mobasher}}]{matthee15}
{Matthee}, J., {Sobral}, D., {Santos}, S., {et~al.} 2015, \mnras, 451, 400

\bibitem[{{Matthee} {et~al.}(2019){Matthee}, {Sobral}, {Boogaard},
  {R{\"o}ttgering}, {Vallini}, {Ferrara}, {Paulino-Afonso}, {Boone},
  {Schaerer}, \& {Mobasher}}]{matthee19}
{Matthee}, J., {Sobral}, D., {Boogaard}, L.~A., {et~al.} 2019, \apj, 881, 124

\bibitem[{{Ota} {et~al.}(2017){Ota}, {Iye}, {Kashikawa}, {Konno}, {Nakata},
  {Totani}, {Kobayashi}, {Fudamoto}, {Seko}, {Toshikawa}, {Ichikawa},
  {Shibuya}, \& {Onoue}}]{ota17}
{Ota}, K., {Iye}, M., {Kashikawa}, N., {et~al.} 2017, \apj, 844, 85

\bibitem[{{Ouchi} {et~al.}(2008){Ouchi}, {Shimasaku}, {Akiyama}, {Simpson},
  {Saito}, {Ueda}, {Furusawa}, {Sekiguchi}, {Yamada}, {Kodama}, {Kashikawa},
  {Okamura}, {Iye}, {Takata}, {Yoshida}, \& {Yoshida}}]{ouchi08}
{Ouchi}, M., {Shimasaku}, K., {Akiyama}, M., {et~al.} 2008, \apjs, 176, 301

\bibitem[{{Ouchi} {et~al.}(2010){Ouchi}, {Shimasaku}, {Furusawa}, {Saito},
  {Yoshida}, {Akiyama}, {Ono}, {Yamada}, {Ota}, \& {Kashikawa}}]{ouchi10}
{Ouchi}, M., {Shimasaku}, K., {Furusawa}, H., {et~al.} 2010, \apj, 723, 869

\bibitem[{{Santos} {et~al.}(2016){Santos}, {Sobral}, \& {Matthee}}]{santos16}
{Santos}, S., {Sobral}, D., \& {Matthee}, J. 2016, \mnras, 463, 1678

\bibitem[{{Sobral} {et~al.}(2015){Sobral}, {Matthee}, {Darvish}, {Schaerer},
  {Mobasher}, {R{\"o}ttgering}, {Santos}, \& {Hemmati}}]{sobral15}
{Sobral}, D., {Matthee}, J., {Darvish}, B., {et~al.} 2015, \apj, 808, 139

\bibitem[{Songaila {et~al.}(2018)Songaila, Hu, Barger, Cowie, Hasinger,
  Rosenwasser, \& Waters}]{songaila18}
Songaila, A., Hu, E.~M., Barger, A.~J., {et~al.} 2018, \apj, 859, 91

\bibitem[{{Zheng} {et~al.}(2017){Zheng}, {Wang}, {Rhoads}, {Infante},
  {Malhotra}, {Hu}, {Walker}, {Jiang}, {Jiang}, {Hibon}, {Gonzalez}, {Kong},
  {Zheng}, {Galaz}, \& {Barrientos}}]{zheng17}
{Zheng}, Z.-Y., {Wang}, J., {Rhoads}, J., {et~al.} 2017, \apjl, 842, L22

\end{thebibliography}

\end{document}